%% file: main.tex
\documentclass[a4paper,11pt]{article}
\usepackage{jheppub} 
\usepackage{lineno}
\usepackage{main-defs}
\usepackage{array}
\usepackage{longtable}
\usepackage{graphicx}
\usepackage{soul}
\usepackage{subcaption}
\usepackage{arydshln}            
\usepackage{changes}
\definechangesauthor[color=teal]{GM}

\title{\boldmath Machine learning with low-level calorimeter signals for beyond the Standard Model photon signatures} 
\title{\boldmath Transformer-based machine learning using low-level calorimeter signals for collimated photon identification at collider experiments} 

\author[a]{Gabriel Matos}
\author[b]{Lauren Larson}
\author[c]{Abhilasha Dave}
\author[a]{Maria Bressan}
\author[b]{Azal Amer}
\author[a]{Cindy Liu}
\author[d]{Nikiforos Nikiforou}
\author[a]{Jonathan Long}
\author[b]{Timothy Andeen}
\author[a]{John Parsons}
\author[c]{Julia Gonski} 
\affiliation[a]{Nevis Laboratories, Columbia University, 136 S Broadway, Irvington, NY 10533, USA}
\affiliation[b]{Department of Physics, The University of Texas at Austin, Austin, TX 78712, USA}
\affiliation[c]{SLAC National Accelerator Laboratory, 2575 Sand Hill Rd, Menlo Park, CA 94025, USA}
\affiliation[d]{CERN, 1211 Geneva 23, Switzerland}

\emailAdd{gabriel.pinheiro.matos@cern.ch}

\abstract{

Electromagnetic calorimeters provide essential information for reconstructing and selecting both Standard Model (SM) and potential beyond the SM physics events at high-energy particle colliders. 
The fine-grained segmentation of modern calorimeters captures rich information about the internal structure of particle showers, much of which is discarded by conventional high-level reconstruction methods. 
In this work, we leverage calorimeter cell-level information to classify highly collimated diphoton signatures, arising from the decay of light axion-like particles, from isolated single-photon showers. 
We systematically compare a range of machine learning architectures, spanning high-level, shower shape variable-based approaches and direct cell-level methods. 
Cell-level machine learning shows significantly superior classification ability, with a Transformer in particular representing the best performance among six different architectures studied, and an MLP Mixer representing a resource-constrained alternative for potential real-time, trigger-level applications. 
Beyond classification, the Transformer model developed enables direct invariant mass regression from calorimeter cells, improving the characterization of light resonances and providing an additional handle in reducing the $\pi^0$ and $\eta$ fake photon backgrounds. 
These results demonstrate that cell-level machine learning methods can extend calorimeter-based particle identification and performance well beyond the capabilities of current conventional techniques.
}

\begin{document}
\maketitle
\flushbottom

\input{sections/intro}
\input{sections/samples}
\input{sections/model}
\input{sections/results}
\input{sections/conclusions}

\acknowledgments
The authors are grateful to Ryan Roberts and Francesco Di Bello for many helpful discussions throughout the development of this paper. GM, CL, JL, and JP are supported by the National Science Foundation under Grant No. PHY-2310080. MB is supported by the National Science Foundation under Grant No. DGE-2437839. LL, AA, and TA are supported by the U.S. Department of Energy under Grant No. DE-SC0007890. AD and JG are supported by the U.S. Department of Energy under Contract No. DE-AC02-76SF00515.

\appendix

\section{Code Availability}
\label{sec:code_availability}

The samples used in this study and relevant machine learning code are publicly available at \href{https://zenodo.org/records/18825682}{zenodo.org/records/18825682} and \href{https://github.com/gabrielpmatos/calo-ml}{github.com/gabrielpmatos/calo-ml}. The trained model weights for the Transformer, including the evaluated test set, are also provided in \href{https://zenodo.org/records/23661362}{zenodo.org/records/23661362}.

\section{Simulated Samples}
\label{sec:sim_samples}

\begin{figure}[h]
    \centering
    \includegraphics[width=0.49\linewidth]{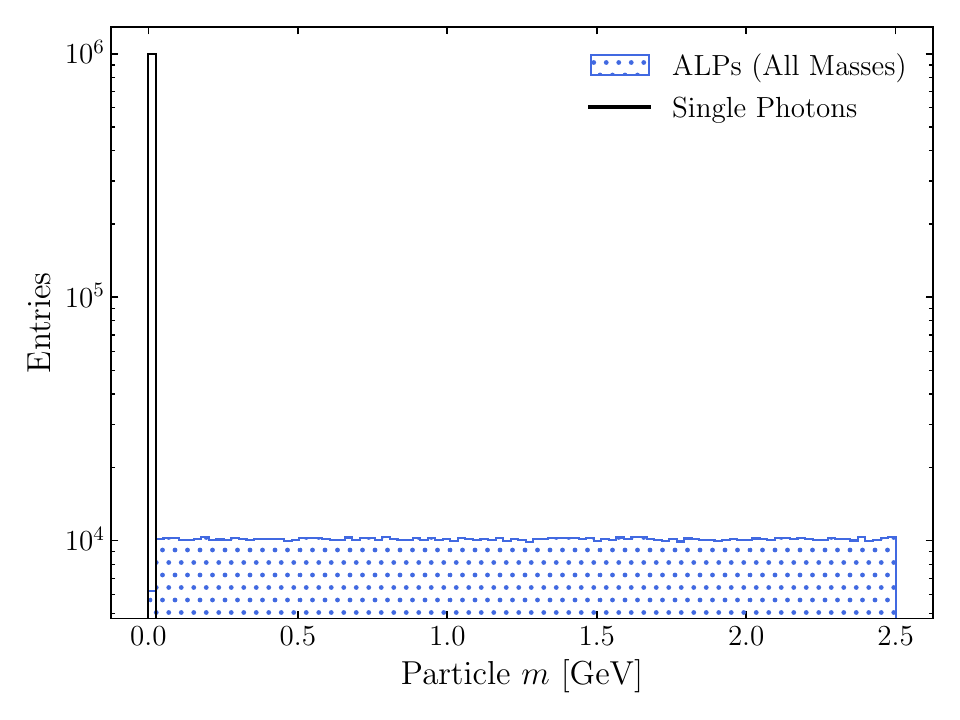}
    \includegraphics[width=0.49\linewidth]{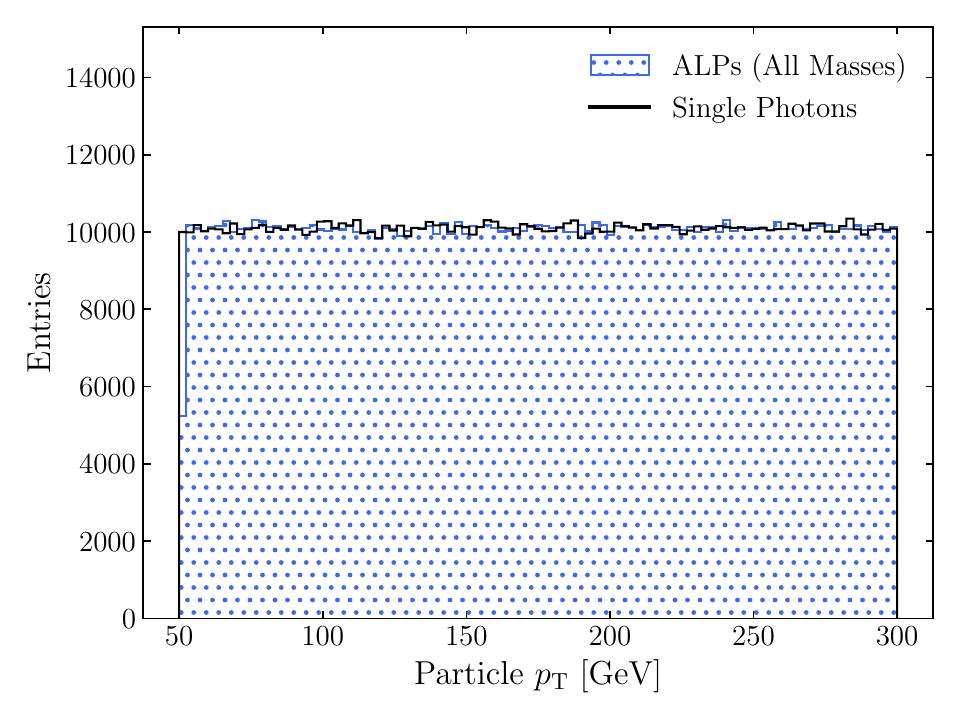}
    \caption{Particle masses (left) and momenta (right) generated in the simulation. ALP masses are generated in the 0.01--2.5~GeV range. Both single photon and ALP momenta are produced in the 50--300~GeV range.}
    \label{fig:simulated_m_and_pt}
\end{figure}

\begin{figure}[h]
    \centering
    \includegraphics[width=0.49\linewidth]{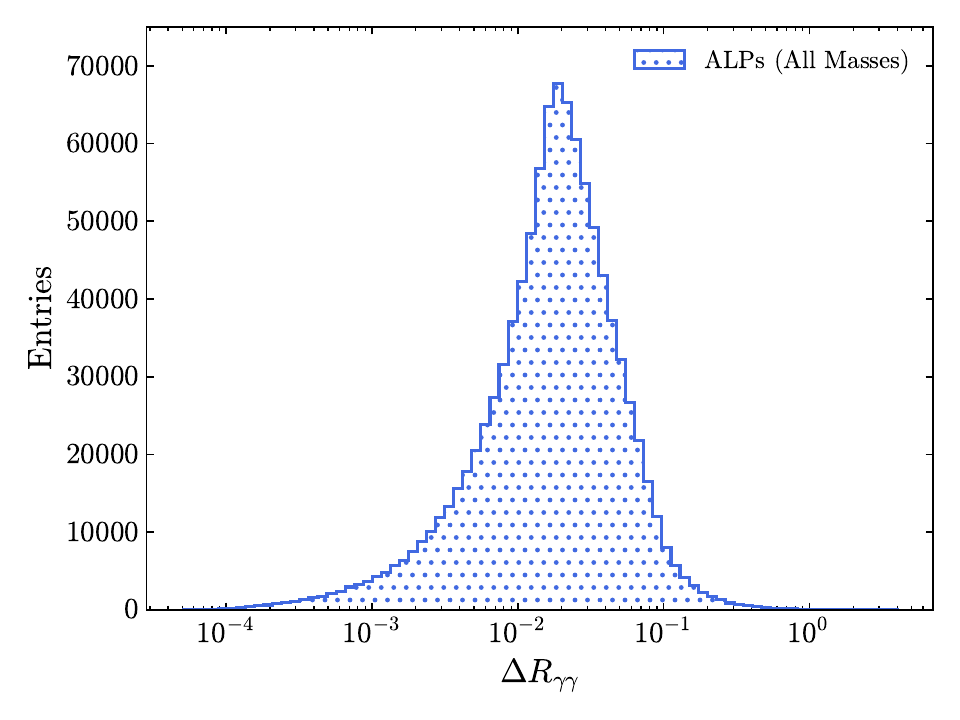}
    \includegraphics[width=0.49\linewidth]{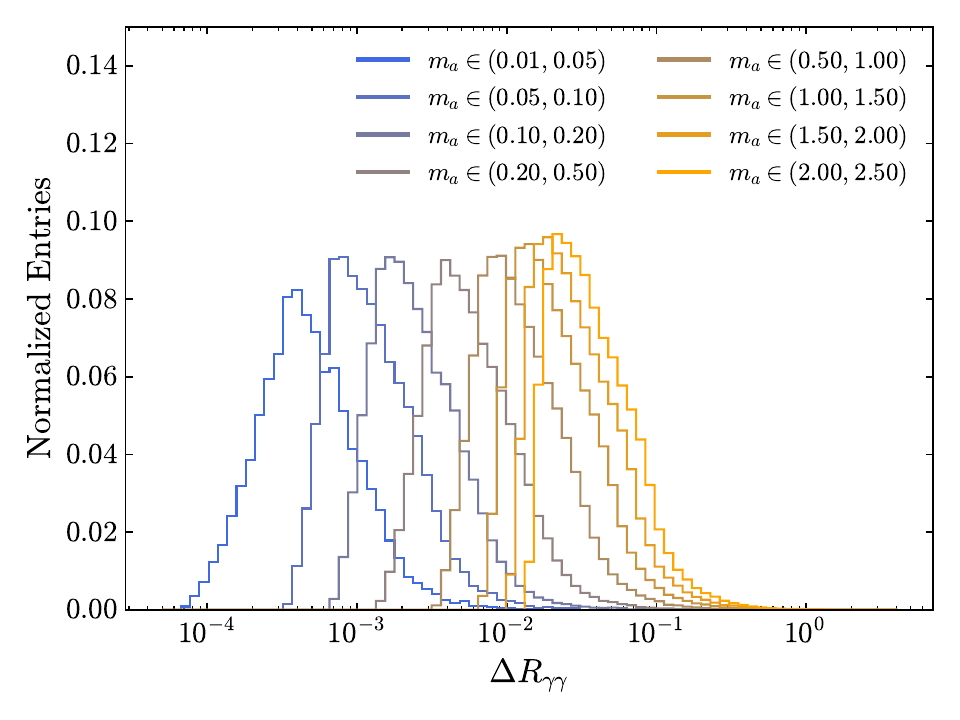}
    \caption{Inclusive $\Delta R_{\gamma\gamma}$ distribution (left) and in bins of ALP mass (right) of the photons from the simulated ALPs. The true opening angles range from approximately $10^{-4}$ to $10^{-1}$.}
    \label{fig:simulated_dr}
\end{figure}

\begin{figure}[h]
    \centering
    \includegraphics[width=0.6\linewidth]{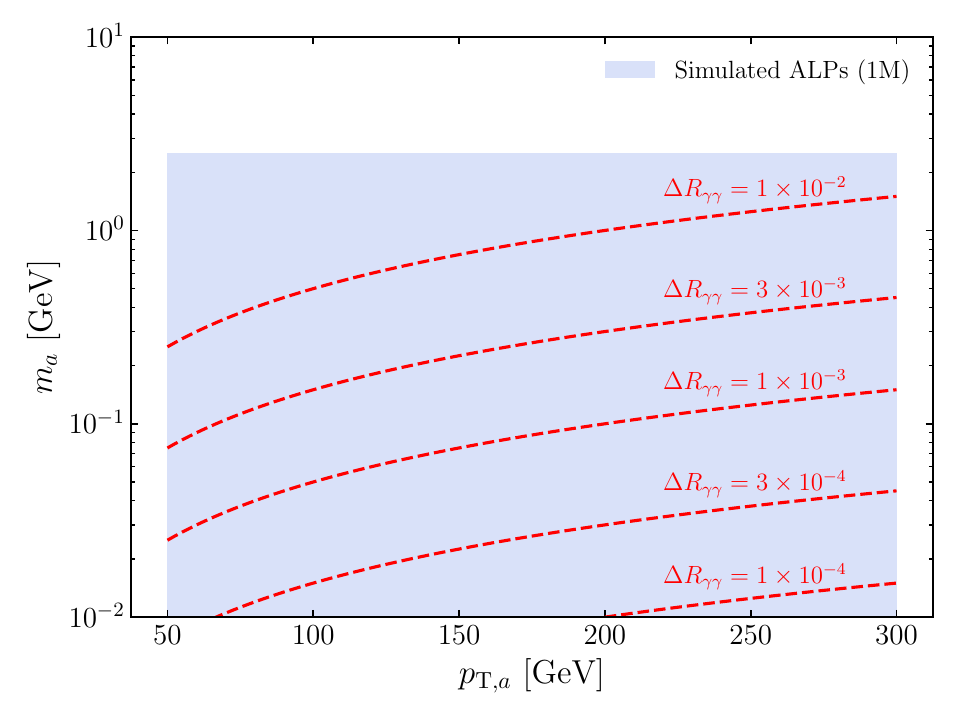}
    \caption{Illustration of simulated ALP samples, where the mass and momentum range shown is sampled uniformly during signal generation. $\Delta R_{\gamma\gamma}$ bands are shown to illustrate different collimation levels as a function of $m_a$ and $p_{\mathrm{T},a}$.}
    \label{fig:alp_grid}
\end{figure}

\begin{figure}[h]
    \centering
    \includegraphics[width=0.6\linewidth]{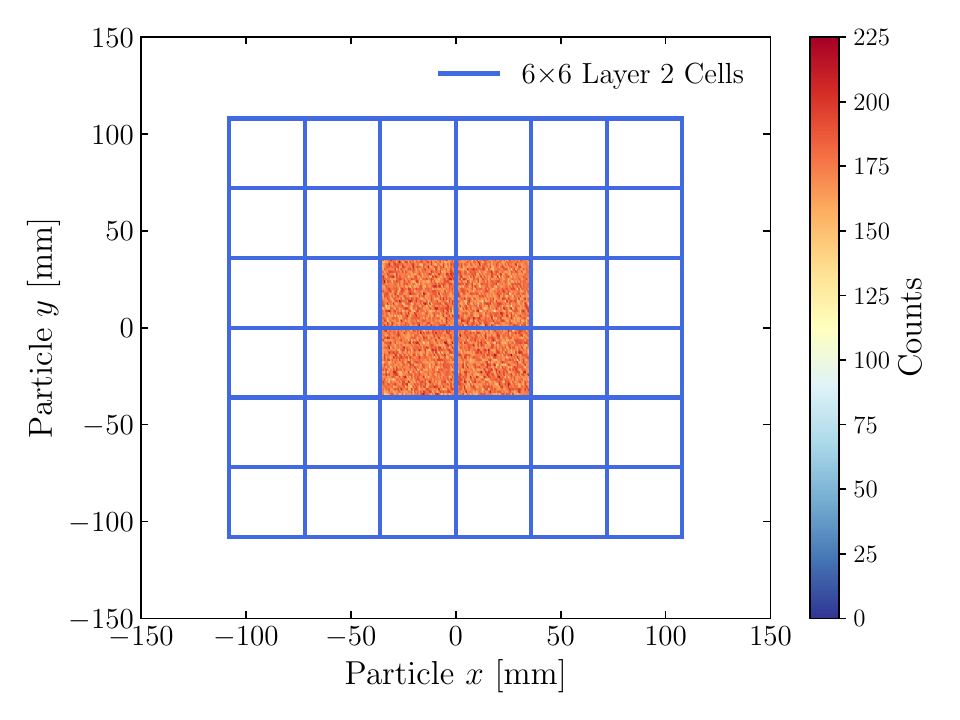}
    \caption{Particle gun smearing in a 2$\times$2 grid of middle layer cells in the signal generation. The incidence positions in $(x,y)$ are histogrammed inclusively of both ALPs and single photons, and a grid of 6$\times$6 middle layer cells is overlaid to illustrate the smearing.}
    \label{fig:smearing}
\end{figure}

Figure~\ref{fig:simulated_m_and_pt} shows the distribution of the simulated masses and momenta for both the ALP and single photon samples, and Figure~\ref{fig:simulated_dr} shows the true $\Delta R_{\gamma\gamma}$ distribution both inclusively and in bins of $m_a$ for the ALPs. The simulated ALP masses are further shown in Figure~\ref{fig:alp_grid}. The $m_a$ and $p_{\mathrm{T},a}$ ranges shown, namely 0.01--2.5 GeV and 50--300 GeV, respectively, are covered uniformly in the signal generation. $\Delta R_{\gamma\gamma}$ bands are drawn, displaying the overall level of collimation as a function of the ALP mass and momentum. The particle gun position (and hence, the particle incidence in the calorimeter) is smeared uniformly across a block of 2$\times$2 middle layer cells to prevent any biases due to the calorimeter geometry. The smearing is shown in Figure~\ref{fig:smearing}, where an overlay grid of 6$\times$6 middle layer cells is drawn, and the number of particle incidences at a given ($x,y$) location is histogrammed.

\section{Shower Shape Variables}
\label{sec:ssvs}

\begin{figure}[tbh]
    \centering

    \begin{subfigure}{0.32\textwidth}
        \centering
        \includegraphics[width=\linewidth]{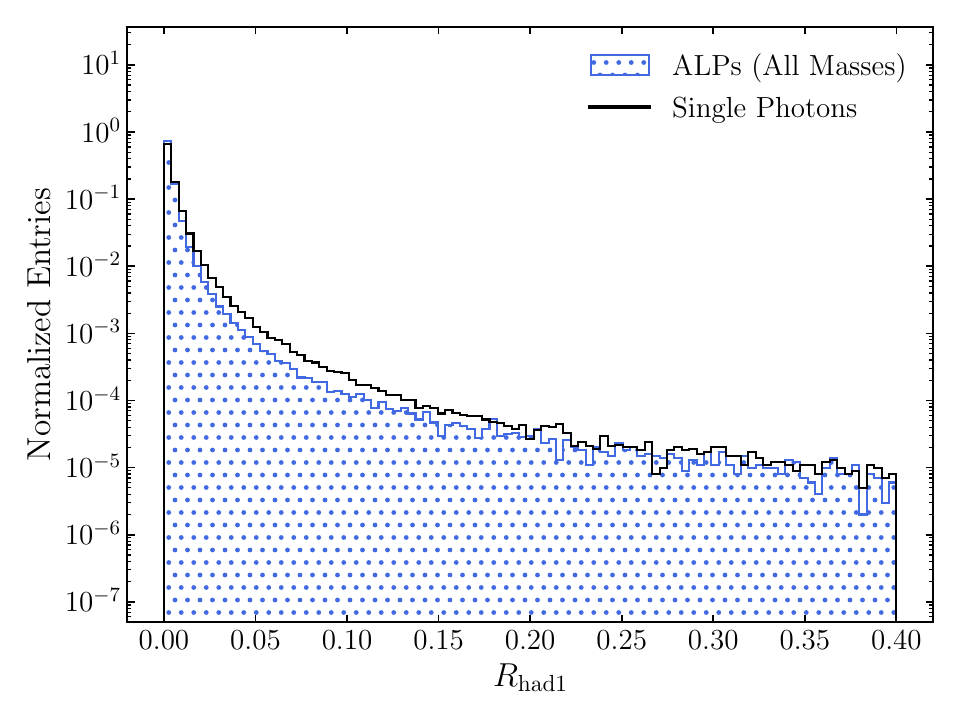}
    \end{subfigure}
    \hfill
    \begin{subfigure}{0.32\textwidth}
        \centering
        \includegraphics[width=\linewidth]{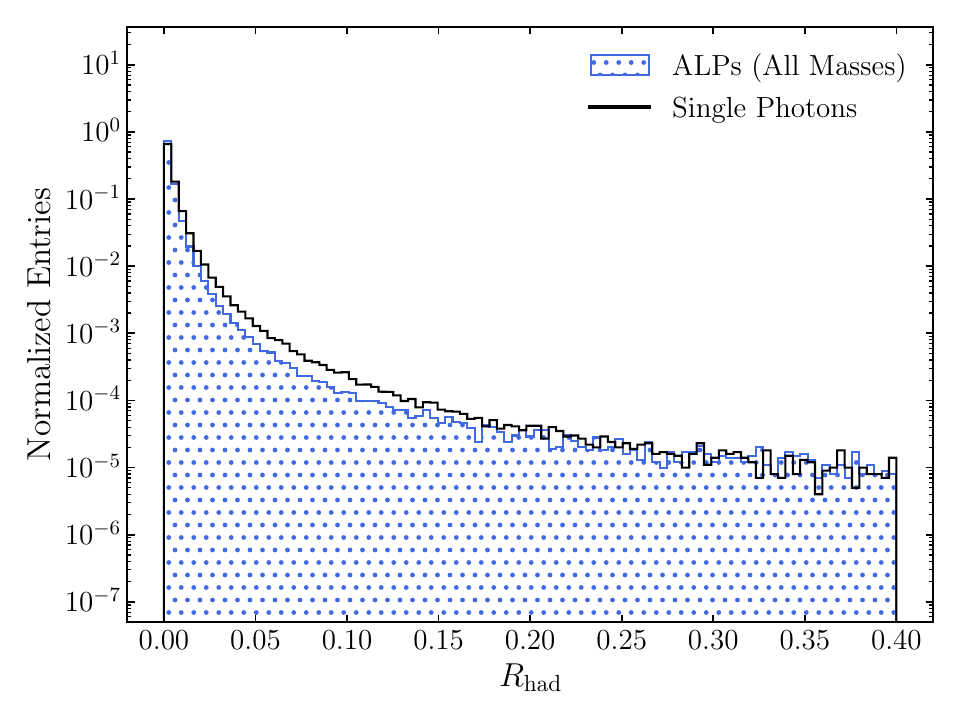}
    \end{subfigure}
    \hfill
    \begin{subfigure}{0.32\textwidth}
        \centering
        \includegraphics[width=\linewidth]{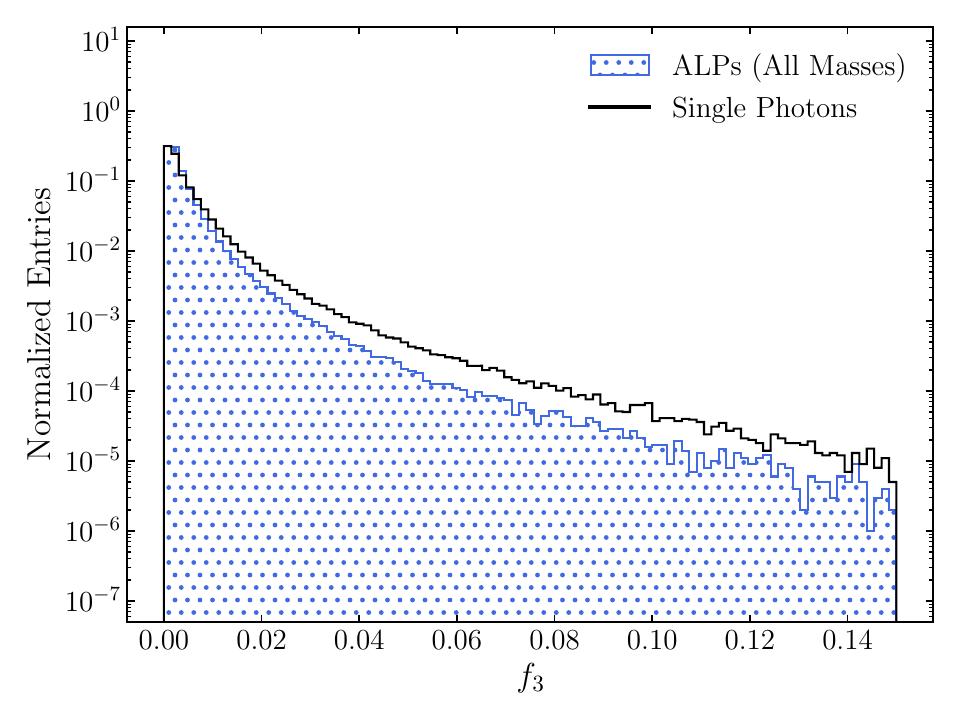}
    \end{subfigure}

    \vspace{0.5em}

    \begin{subfigure}{0.32\textwidth}
        \centering
        \includegraphics[width=\linewidth]{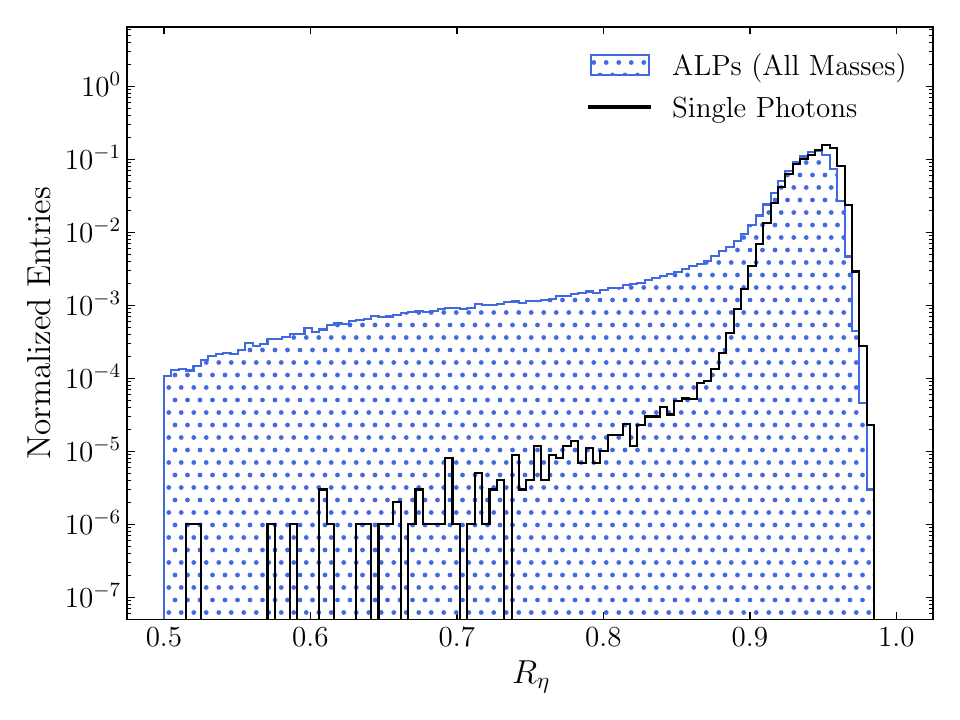}
    \end{subfigure}
    \hfill
    \begin{subfigure}{0.32\textwidth}
        \centering
        \includegraphics[width=\linewidth]{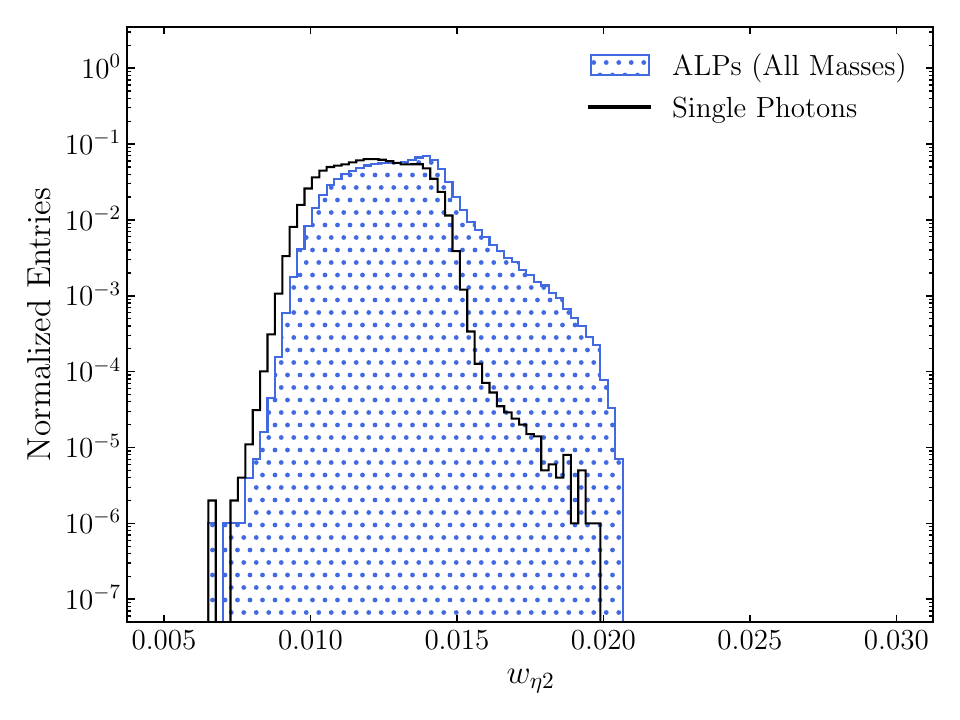}
    \end{subfigure}
    \hfill
    \begin{subfigure}{0.32\textwidth}
        \centering
        \includegraphics[width=\linewidth]{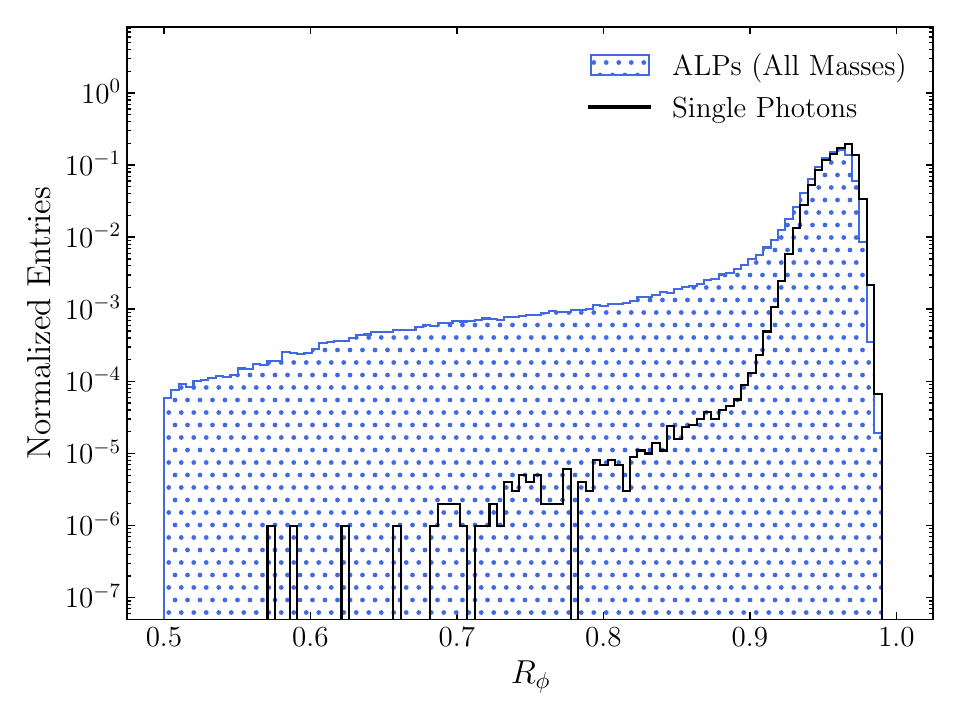}
    \end{subfigure}

    \vspace{0.5em}

    \begin{subfigure}{0.32\textwidth}
        \centering
        \includegraphics[width=\linewidth]{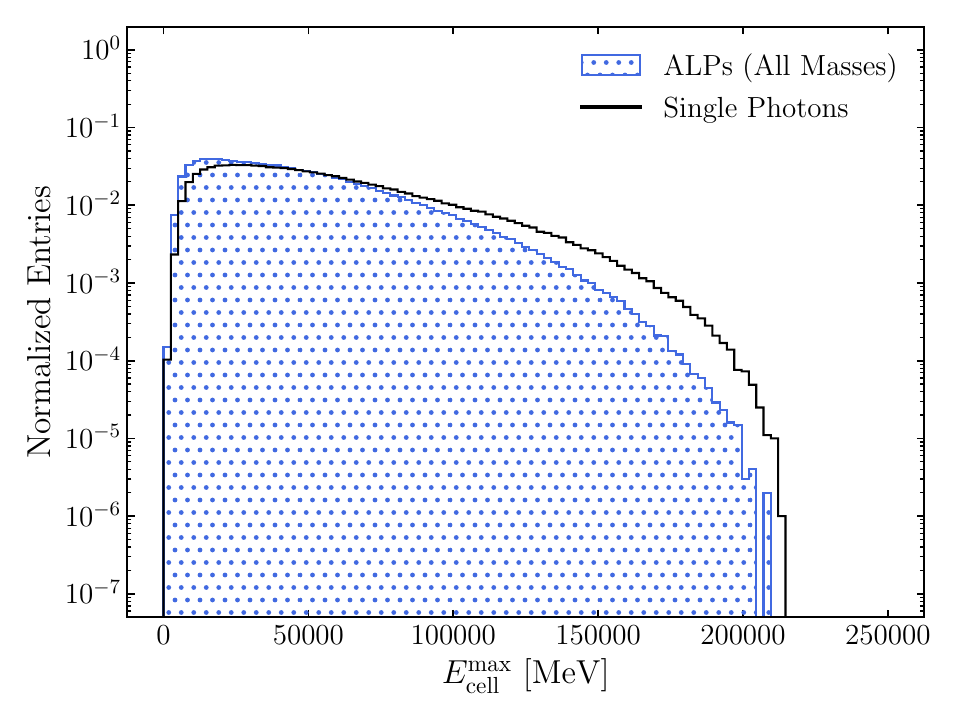}
    \end{subfigure}
    \hfill
    \begin{subfigure}{0.32\textwidth}
        \centering
        \includegraphics[width=\linewidth]{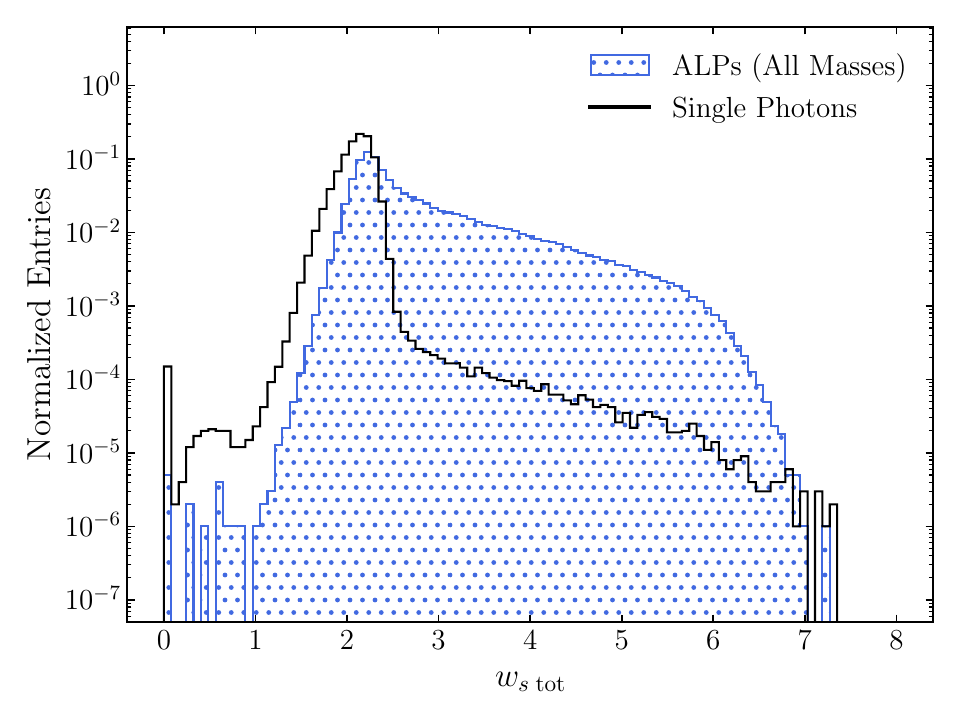}
    \end{subfigure}
    \hfill
    \begin{subfigure}{0.32\textwidth}
        \centering
        \includegraphics[width=\linewidth]{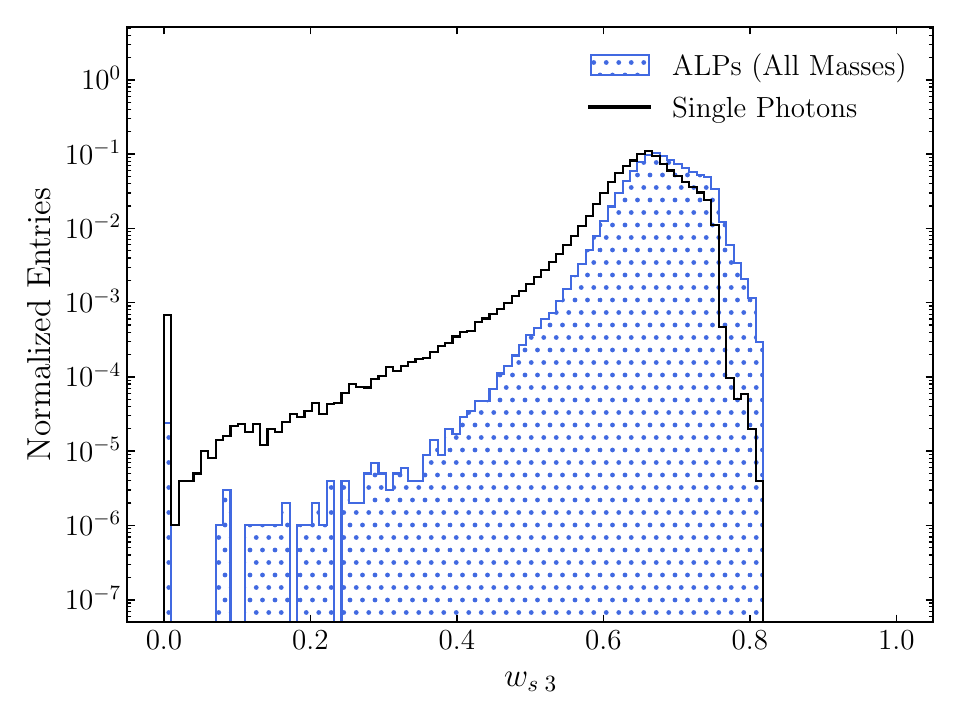}
    \end{subfigure}

    \vspace{0.5em}

    \begin{subfigure}{0.32\textwidth}
        \centering
        \includegraphics[width=\linewidth]{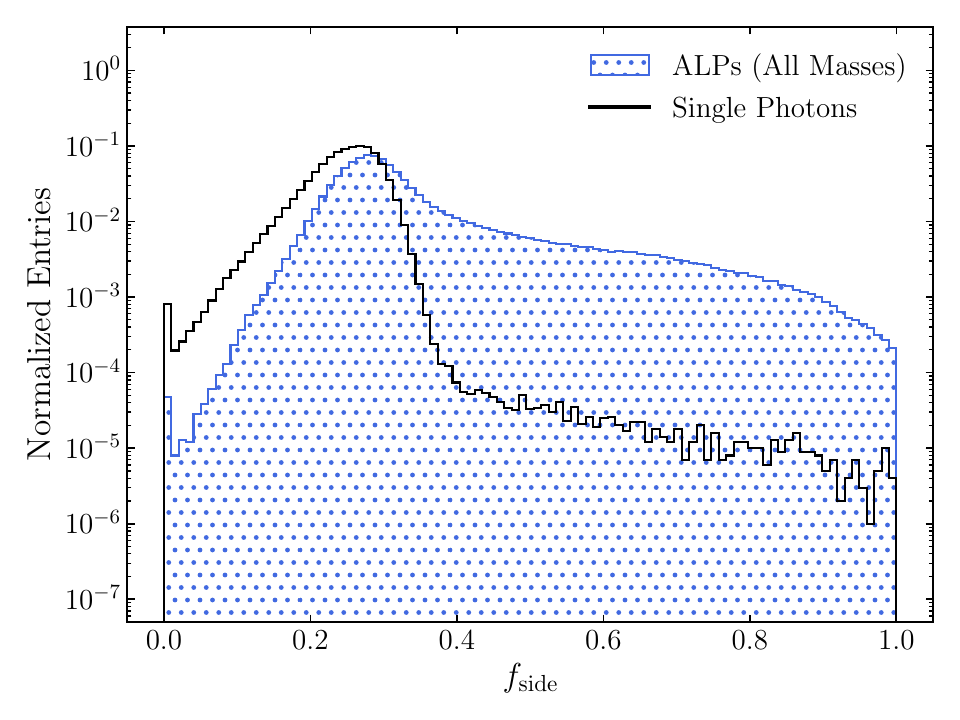}
    \end{subfigure}
    \hfill
    \begin{subfigure}{0.32\textwidth}
        \centering
        \includegraphics[width=\linewidth]{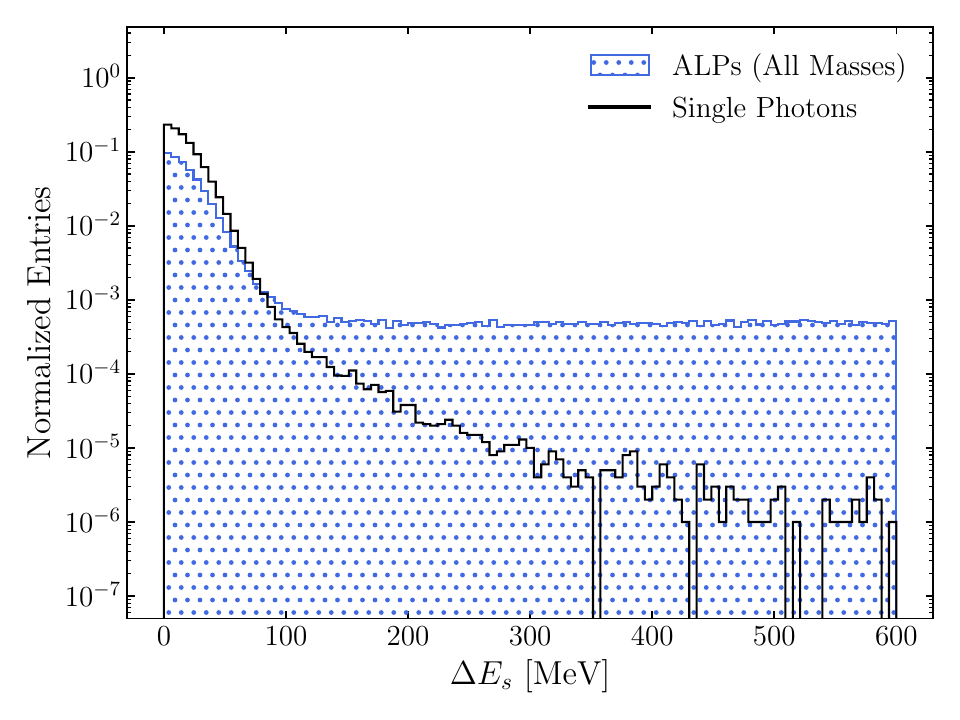}
    \end{subfigure}
    \hfill
    \begin{subfigure}{0.32\textwidth}
        \centering
        \includegraphics[width=\linewidth]{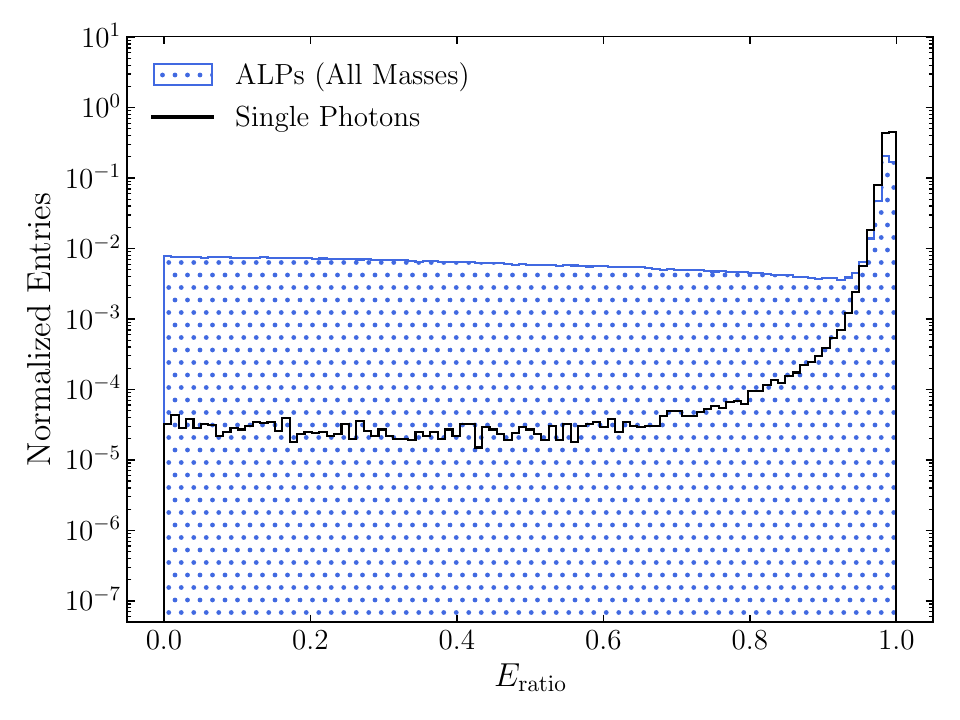}
    \end{subfigure}

    \vspace{0.5em}

    \begin{subfigure}{0.5\textwidth}
        \centering
        \includegraphics[width=0.64\linewidth]{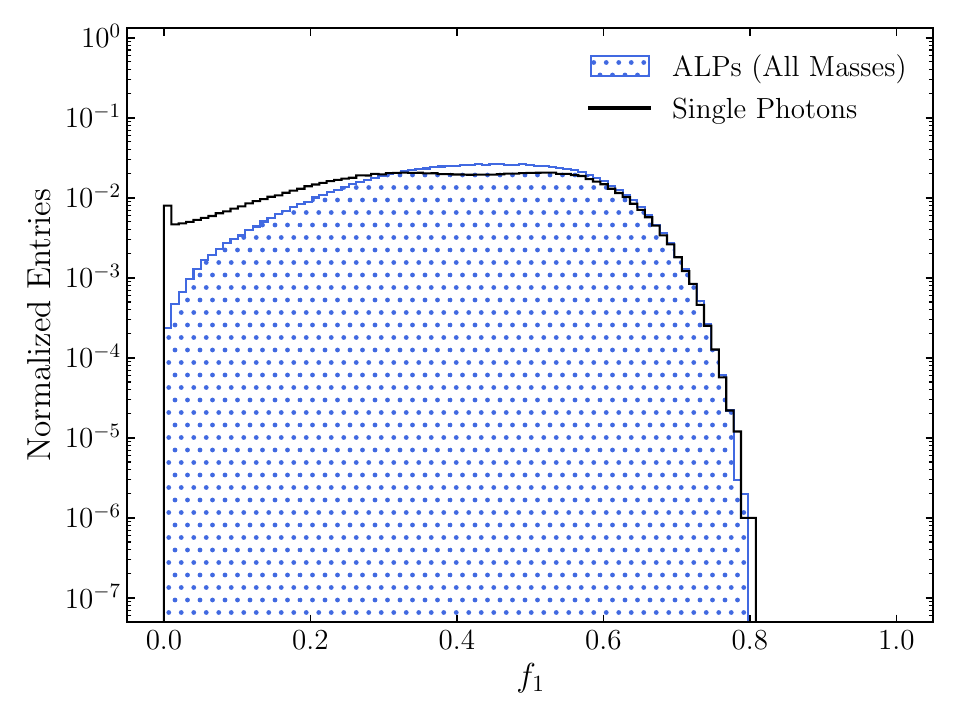}
    \end{subfigure}

    \caption{Distribution of the SSVs used for the BDT and DNN trainings. The ALP samples are shown inclusively for all masses and momenta generated. The single photon sample is shown for all momenta.}
    \label{fig:ssvs}
\end{figure}

The definitions of the SSVs used in this study are identical to those used for photon identification in ATLAS \cite{ATLAS:2016xha}, with the addition of a variable describing the energy of the maximum energy cell in the middle layer of the EM calorimeter, $E_{\mathrm{cell}}^{\mathrm{max}}$. Figure~\ref{fig:ssvs} shows the SSV distributions for the ALP sample (displayed inclusively for all masses and momenta generated) and single photon sample.

\section{Decay Energy Asymmetry Impact on Transformer Performance}
\label{sec:asymmetry}

\begin{figure}[h]
    \centering
    \includegraphics[width=0.49\linewidth]{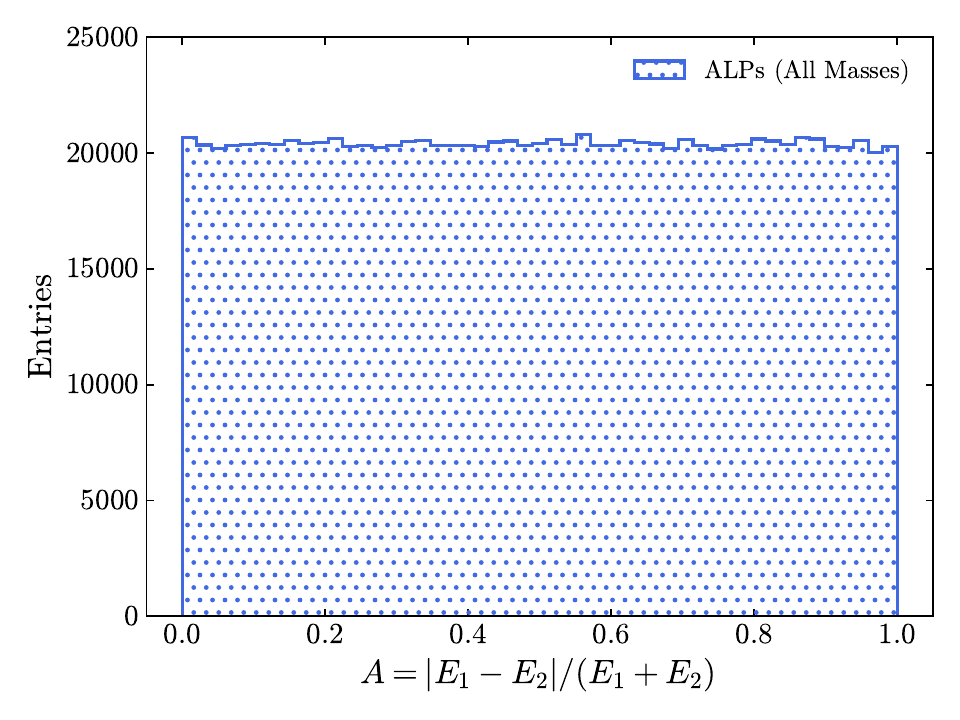}
    \includegraphics[width=0.49\linewidth]{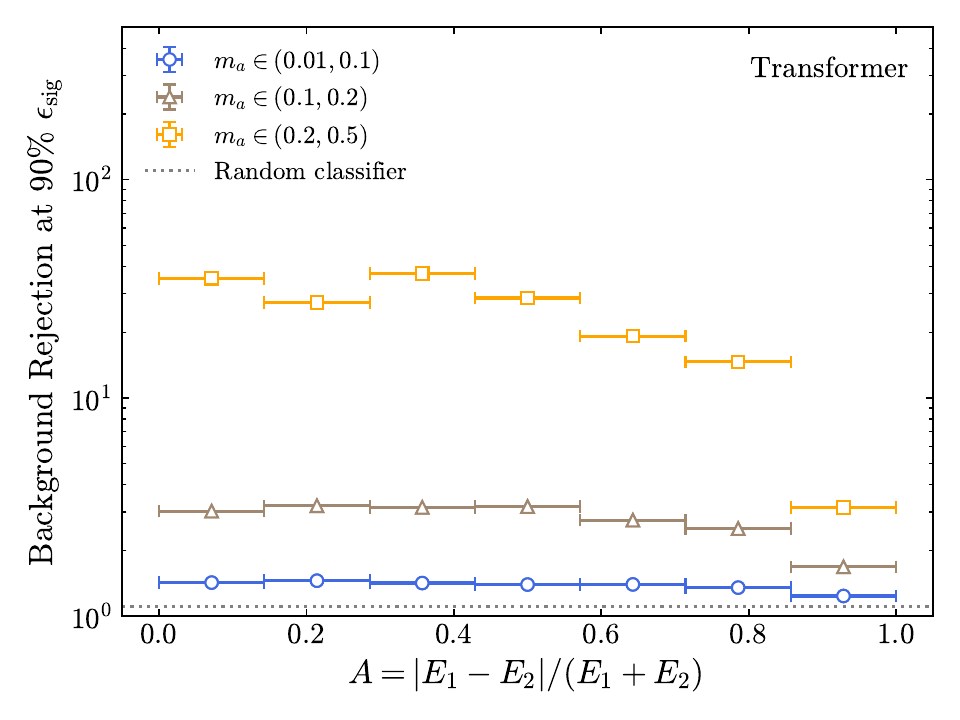}
    \caption{Simulated decay energy asymmetry (left) and the Transformer's performance as a function of this asymmetry (right), shown in terms of the background rejection at 90\% signal efficiency for the three lowest ALP mass bins. The asymmetry distribution is uniform, and the performance is roughly independent of the asymmetry except in the highest asymmetry bin, where there is a degradation for very asymmetric decays.}
    \label{fig:asymmetry}
\end{figure}

Asymmetric decays of the ALPs can have a larger angular separation, producing a hard first photon and a soft second photon. A study was done to assess the impact of the decay energy asymmetry on the Transformer performance, in particular in the lowest ALP mass bins where the classification is most challenging. The decay energy asymmetry can be characterized by

\begin{equation}
    A = \frac{|E_1 - E_2|}{E_1 + E_2},
\end{equation}

\noindent where $E_1$ and $E_2$ represent the leading and subleading photon energies from the $a\rightarrow \gamma\gamma$ decay. Figure~\ref{fig:asymmetry} shows both the generated asymmetry distribution from the simulated ALPs, as well as the  Transformer background rejection at 90\% signal efficiency in the three lowest ALP mass bins as a function of the asymmetry. Because the ALP decays are simulated to be isotropic in the ALP's rest frame, the true asymmetry distribution is uniform. The Transformer performance is also mostly independent of the decay asymmetry, except in the highest asymmetry bin where the performance degrades.

\clearpage
\bibliographystyle{JHEP}
\bibliography{biblio.bib}


\end{document}

%% file: sections/intro.tex
\section{Introduction}
\label{sec:intro}



The Standard Model (SM) of particle physics leaves several key fundamental questions unresolved, including the nature of dark matter and the origin of the observed matter–antimatter asymmetry in the universe. 
Addressing these questions requires new physics beyond the SM (BSM), whose signatures have so far remained experimentally elusive and continue to motivate the development of novel search strategies in collider physics experiments. 
One such strategy is to exploit low-level detector signals, which provide a more detailed representation of collision events than standard methods based on high-level reconstructed objects. 
Machine learning (ML) is key to exploiting the high-dimensional correlations present in these low-level signals, improving sensitivity to subtle BSM signatures while introducing challenges in complex data modeling and processing. 




This work explores the capability of such an approach in calorimetry. In a fine-grained electromagnetic (EM) calorimeter, such as the liquid argon (LAr) calorimeter~\cite{CERN-LHCC-96-041} of the ATLAS Experiment~\cite{atlasdetectorpaper} at the CERN Large Hadron Collider (LHC)~\cite{Evans:2008zzb}, the smallest detector units are the calorimeter cells. 
Rather than relying on standard reconstruction techniques, which aggregate across the calorimeter cells and can obscure key cell-level correlations, the signals in the individual cells can be leveraged directly as a high-dimensional input for a more nuanced probe of EM shower energy deposition. 

\begin{figure}[tbh]
\centering
\includegraphics[width=1\textwidth]{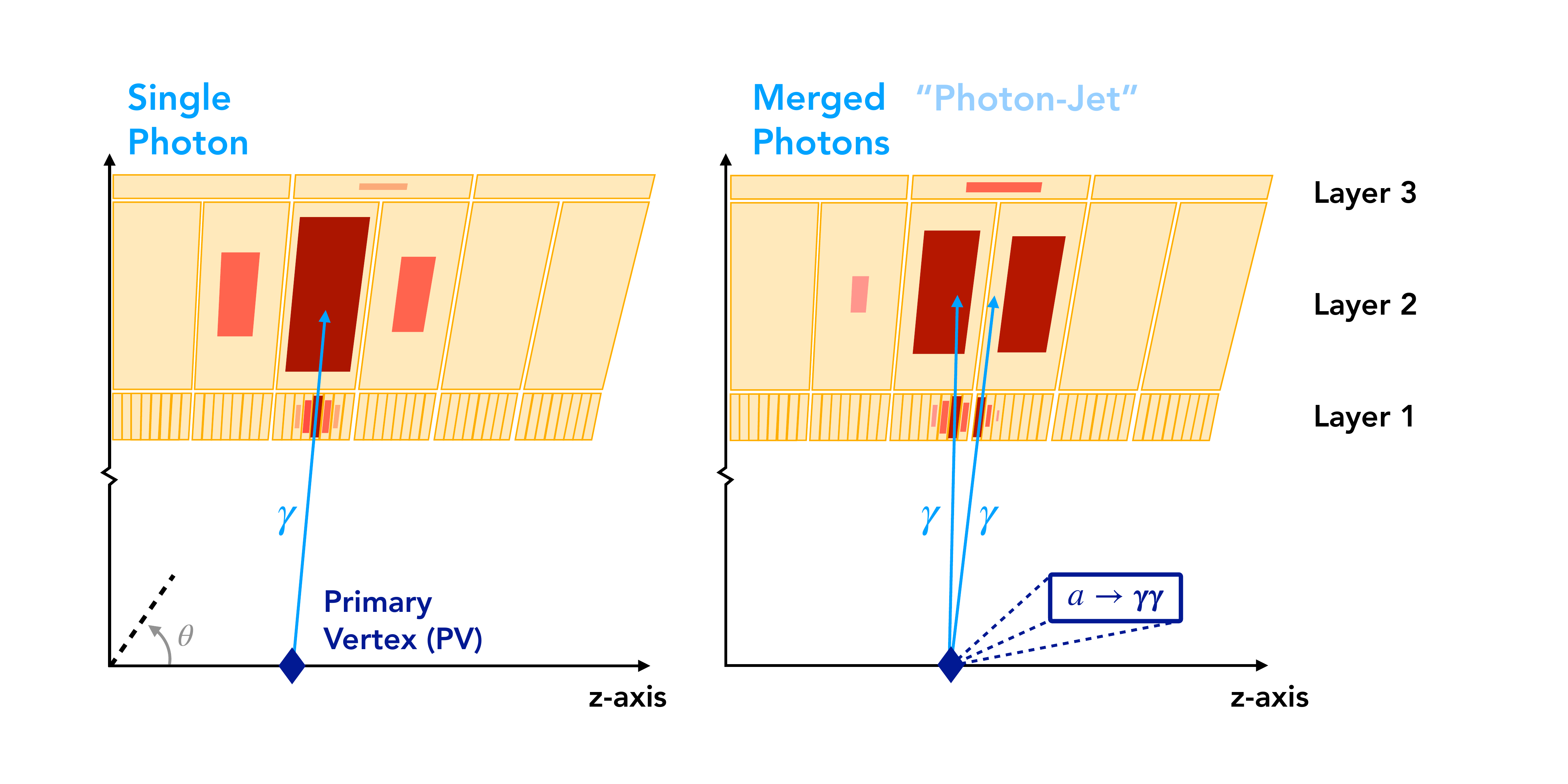}
\caption{\label{fig:cartoon} Schematic diagram of different shower topologies in an ATLAS-like EM calorimeter due to (left) a standard EM shower arising from a single photon, and (right) an overlapping diphoton ``photon-jet'' originating from a light and highly boosted ALP, $a$, decaying into two photons.}
\end{figure}

The subtle correlations within an EM shower are particularly important for distinguishing isolated photons from light, highly boosted particles that immediately decay into multiple photons which are so collimated they cannot be resolved as separate objects, forming ``photon-jets"~\cite{Dasgupta_2016,Aparicio_2016,Knapen_2016,Chang_2016}. 
Such signatures arise in SM meson decays (e.g. $\pi^0\rightarrow \gamma\gamma$ and $\eta\rightarrow\gamma\gamma$), and in a variety of BSM scenarios. 
In this study, we focus on the classification of photon-jets originating from the decay of light axion-like particles (ALPs)~\cite{PhysRevD.16.1791,PhysRevLett.38.1440,PhysRevLett.40.279,PhysRevLett.40.223,Mimasu:2014nea,Bauer_2017,Agrawal_2021} to two photons. 
ALPs are a well-motivated dark matter candidate~\cite{Yan:2025alk,Lentz:2026bmp,Akgumus:2025mrh,Chadha-Day:2021szb,Marsh:2015xka,Dine:1982ah,Abbott:1982af,Preskill:1982cy} that, for ALP masses in the $\mathcal{O}$(10--100) MeV range, can exhibit a significant branching ratio to the diphoton final state ~\cite{Bauer_2017, Dasgupta_2016, Jaeckel_2016,ellwanger2016750gevdiphotonsignal,Agrawal_2021}. These models often predict couplings between the ALP and the Higgs boson or another heavy scalar mediator, leading to highly boosted ALPs which can manifest as a photon-jet topology in the calorimeters via their $\gamma\gamma$ decays.
Improving the classification of photon-jets can enhance the sensitivity to these ALP signatures, while also improving the ability to reject $\pi^0$ fakes in standard photon identification. 
Figure~\ref{fig:cartoon} provides a schematic diagram illustrating the detector response from an EM shower resulting from an isolated single photon and from a photon-jet originating in an ALP decay.

 
Previous approaches to photon-jet classification have relied on shower-shape variables (SSVs), which aggregate cell signals across the calorimeter and calculate moments of the EM showers that are traditionally used to reject ``fake" photons due to $\pi^0$ meson decays \cite{photonidcms_2015, ATLAS:2016xha, Sirunyan_2021, photonidatlas_2019}. 
These variables have been incorporated into cut-based and multivariate approaches, including boosted decision trees (BDTs) and deep neural networks (DNNs), for particle identification and photon-jet classification in various contexts~\cite{ATL-PHYS-PUB-2022-022,PhysRevD.99.012008,haa4y_ATLAS}. 
More recent methods have moved toward lower-level, image-based representations of the calorimeter with DNNs~\cite{deOliveira:2018hva} or, more often, convolutional neural networks (CNNs)~\cite{ATL-PHYS-PUB-2023-001, PhysRevLett.134.041801, PhysRevLett.131.101801, Tumasyan_2023, cmscollaboration2026searchexotichiggsboson, Simkina:2023bjm}. 
While these methods improve upon SSV-based approaches, they often rely on mapping detector data onto fixed grids. In non-uniform calorimeters like the ATLAS LAr calorimeter, this often necessitates aggregating across detector layers or region-specific handling, limiting the usable information for classification.
Deep Sets architectures like Particle Flow Networks (PFNs)~\cite{Komiske_2019, ai2024detectinghighlycollimatedphotonjets} address some of these limitations by processing detector elements as permutation-invariant point clouds, enabling flexible cell-level inputs. Such models, however, rely on global pooling operations that can miss important pairwise correlations needed to resolve the photon-jet structure.

This work investigates the use of full calorimeter cell-level granularity to classify photon-jets using a Transformer architecture, which represents the state-of-the-art for modeling high-dimensional, unordered input spaces~\cite{vaswani2023attentionneed}. 
Originally developed for natural language processing, Transformers employ self-attention mechanisms to model pairwise and higher-order interactions between input tokens. Unlike convolutional or recurrent architectures, they naturally accommodate variable-sized inputs and can be designed to be permutation-invariant, making them well-suited for particle physics applications involving unordered detector hits or energy deposits. Transformers have demonstrated strong performance across a wide range of domains in high-energy physics, such as jet flavor tagging in \mbox{ATLAS}~\cite{ATL-PHYS-PUB-2022-027, atlascollaboration2025transformingjetflavourtagging}, or clustering of overlapping EM showers in CMS~\cite{Maidannyk_2026, CMS-DP-2026-135}. 
By explicitly modeling correlations between calorimeter cells while accommodating a variable detector geometry, these architectures provide a flexible framework for learning the detailed shower substructure associated with highly collimated photon decays.


Despite their expressive power, Transformers can be computationally demanding due to the $\mathcal{O}(N^2)$ scaling of attention mechanisms with input size. This consideration is particularly relevant for potential real-time applications in collider experiments, such as trigger-level event selection, where strict latency and resource constraints must be satisfied. To address this, the multi-layer perceptron (MLP) Mixer architecture~\cite{tolstikhin2021mlpmixerallmlparchitecturevision} is also investigated in this study. The MLP Mixer replaces attention with alternating “mixing” operations across input tokens and feature channels, enabling the modeling of global correlations while maintaining a simpler and more hardware-efficient structure. While such models may exhibit reduced classification performance relative to Transformers, they offer advantages in computational efficiency, making them well-suited for deployment in resource-constrained environments, such as field-programmable gate arrays (FPGAs) used in future high-throughput data acquisition systems~\cite{CERN-LHCC-2017-020}.

Through a systematic comparison of Transformer- and MLP Mixer-based approaches for cell-level calorimeter learning against previously leveraged ML architectures, this work provides promising results for the event selection performance in challenging regions of BSM parameter space that can be used in future collider searches, while also expanding the application scope to include potential real-time applications. Such trigger-level implementations could ultimately enable lower trigger thresholds for exotic signatures and broaden sensitivity to BSM physics that may be inaccessible with current online selections.

%% file: sections/samples.tex
\section{Detector Model and Simulated Samples}
\label{sec:samples}

\begin{figure}[h!]
    \centering
    \includegraphics[width=0.85\linewidth]{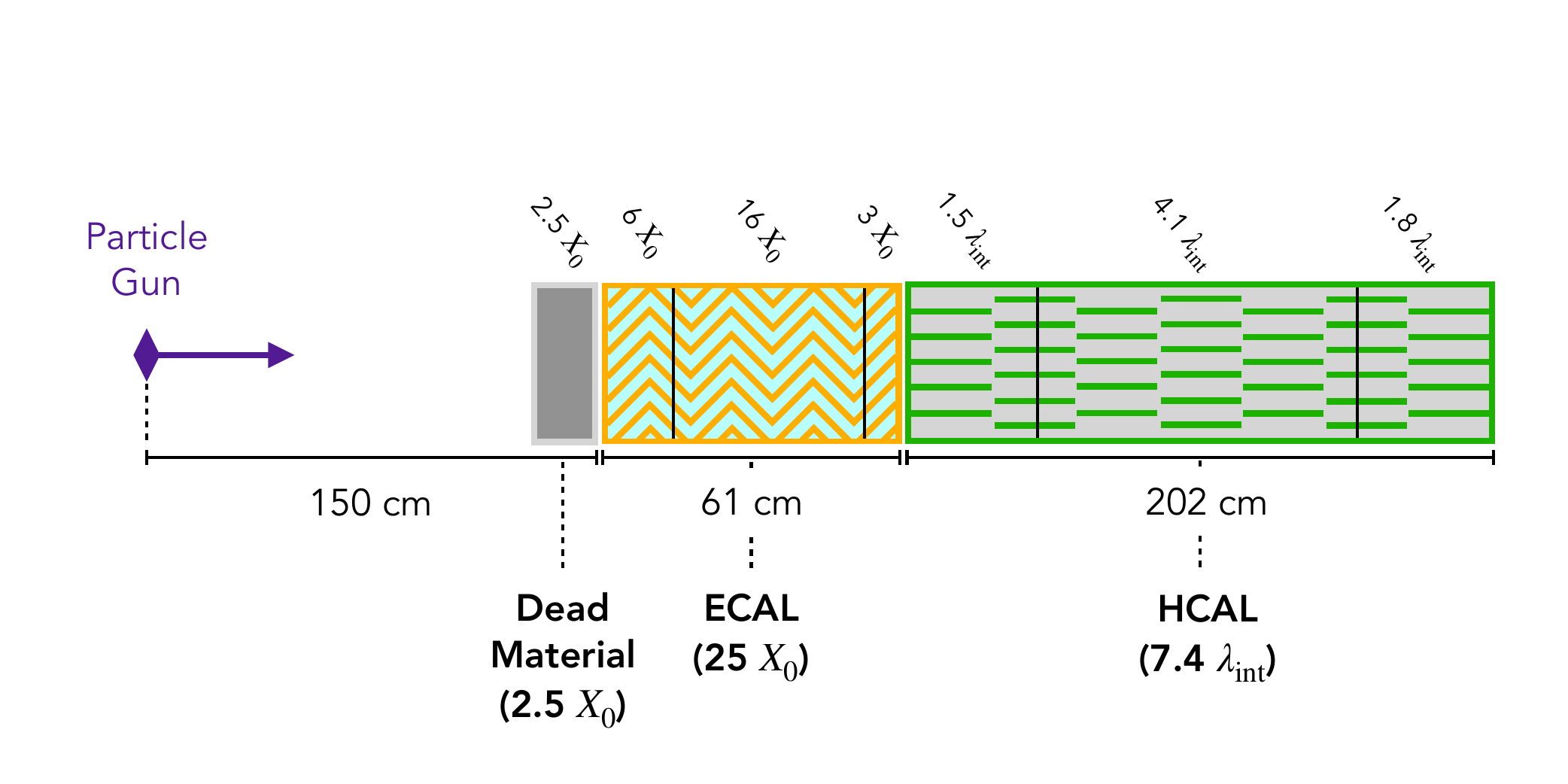}
    \caption{Schematic of the longitudinal distribution and material budget of the ATLAS-like calorimeter used in the simulation. The separation between the different layers is shown by the black lines in the EM and hadronic calorimeters. The sizes of each component are not shown to scale.}
    \label{fig:caloDiagram}
\end{figure}

The cell-level calorimeter response to EM showers from both single photons and photon-jets is simulated using \textsc{Geant4}~\cite{Agostinelli:2002hh}. In the simulation, the \texttt{QGSP\_FTFP\_BERT} physics list is used. The default production range cut of $0.7~\mathrm{mm}$ is applied, and no user step-size limits are imposed. The simulation employs a model~\cite{simplifiedatlassim} designed to mimic the ATLAS calorimeters~\cite{atlasdetectorpaper}, including simplified representations of the accordion geometry of the LAr EM calorimeters and of the scintillating tiles of the Tile hadronic calorimeter. A side view of the calorimeter setup, including the material budget used, is shown in Figure~\ref{fig:caloDiagram}. The model implements the segmentation of the ATLAS calorimeters at pseudorapidity\footnote{As usual for collider experiments, we use a cylindrical coordinate system with its origin at the interaction point
in the center of the detector and the \(z\)-axis along the beam pipe.
The \(y\)-axis points vertically upwards, leaving the \(x\)-axis to be defined in order to provide a right-handed coordinate system. 
Polar coordinates \((r,\phi)\) are used in the transverse plane, 
\(\phi\) being the azimuthal angle around the \(z\)-axis.
The pseudorapidity is defined in terms of the polar angle \(\theta\) as \(\eta = -\ln [\tan(\theta/2)]\).} $\eta=0$. In particular, the EM calorimeter matches the cell geometry of the ATLAS EM barrel (EMB), which includes three longitudinal layers of varying transverse granularity. The middle layer, where the bulk of the EM shower energy is deposited, has roughly square cells with a transverse segmentation corresponding to $\Delta\eta\times\Delta\phi = 0.025\times0.025$. The first layer, which was optimized in ATLAS for the separation of photons from $\pi^0$ mesons, employs cells which are narrow ``strips", eight times finer in $\eta$ and four times coarser in $\phi$ than the middle layer cells, corresponding to $\Delta\eta\times\Delta\phi \approx 0.0031\times 0.1$. The third layer, which primarily serves as a tail catcher for high energy EM showers, has the same $\phi$ granularity as the middle layer but is twice as coarse in $\eta$, corresponding to $\Delta\eta\times\Delta\phi = 0.05 \times 0.025$. The hadronic calorimeter includes additionally three longitudinal layers, for a total of six layers. The face of the calorimeter is modeled as flat, with an extent including an $x\times y$ block of $64\times64$ EM middle layer cells, sufficient to contain the EM showers being simulated. The geometry parameters of the model are summarized in Table~\ref{tab:caloGeo}.

\begin{table}[]
    \centering
    \begin{tabular}{|c|c|c|c|c|}
        \hline
        Layer & $\Delta x\times \Delta y$ [mm$^2$] & $\Delta 
        \eta\times\Delta\phi$ & Total $X_0$\\
        \hline 
        ECAL 1 & 4.5 $\times$ 144 & 0.0031 $\times$ 0.1 & 6 \\
        ECAL 2 & 36 $\times$ 36 & 0.025 $\times$ 0.025 & 16 \\
        ECAL 3 & 72 $\times$ 36 & 0.05 $\times$ 0.025 & 3 \\
        \hline
         & & & Total $\lambda_{\mathrm{int}}$ \\
        \hline
        HCAL 1 & 144 $\times$ 144 & 0.1 $\times$ 0.1 & 1.5 \\
        HCAL 2 & 144 $\times$ 144 & 0.1 $\times$ 0.1 & 4.1 \\
        HCAL 3 & 288 $\times$ 288 & 0.2 $\times$ 0.2 & 1.8 \\
        \hline
        
    \end{tabular}
    \caption{Summary of the geometry of the ATLAS-like calorimeter used in the simulation. The cell dimensions are shown in $\Delta x\times \Delta y$ (where $x$ and $y$ are transverse to the incident particle trajectory) and $\Delta\eta \times \Delta \phi$, by layer. The depth of the EM calorimeter is summarized in terms of radiation lengths ($X_0$), and for the hadronic calorimeter in terms of nuclear interaction lengths ($\lambda_{\mathrm{int}}$).}
    \label{tab:caloGeo}
\end{table}

The ATLAS detector includes a LAr-based presampler in front of the EMB calorimeter, which is used to correct for upstream energy losses in the approximately one and a half radiation lengths ($\approx1.5X_0$) of material at $\eta = 0$. 
This material includes the beampipe, the extensive inner tracking detector (ID), the inner walls of the barrel cryostat, and the superconducting solenoid used to provide a $ \approx 2~\mathrm{T}$ magnetic field for charged particle tracking throughout 
the volume of the ID~\cite{atlasdetectorpaper}. 
The simplified detector model used in this simulation does not include these elements, nor the LAr presampler itself, which contributes an additional $\approx1X_0$ of material before the active LAr EMB accordion calorimeter.    
To approximately account for the impact on EM shower development of this ``dead material'' in front of the calorimeter, the simulation geometry includes a block of aluminum of thickness $2.5X_0$ placed directly in front of the EM calorimeter, as depicted in Figure~\ref{fig:caloDiagram}.

\begin{figure}[tbh]
\centering
\includegraphics[width=0.85\textwidth]{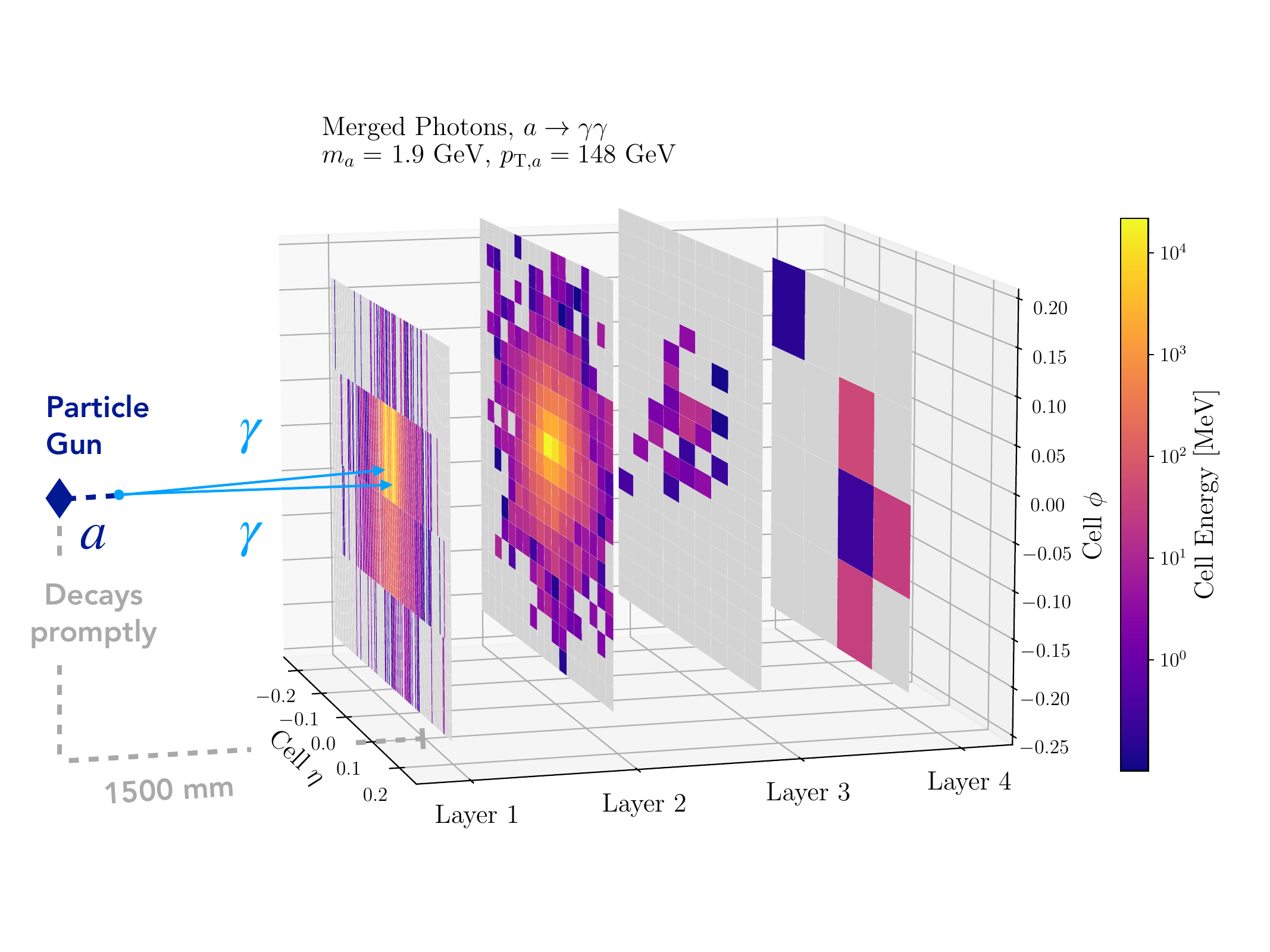}
\caption{\label{fig:caloSim} Visualization of an EM shower resulting from an ALP of \mbox{$p_{\mathrm{T},a} = 148$~GeV} in the calorimeter simulation used in this work. Only the three layers of the EM calorimeter and the first layer of the hadronic calorimeter are shown, but all six calorimeter layers are used in this study. Note that, while the transverse segmentation of the longitudinal layers is illustrated, the depths of the layers are not shown to scale.  The color scale indicates the energy deposited in each calorimeter cell.} 
\end{figure}

Events are generated with a particle gun pointing perpendicularly to the front face of the calorimeter, producing either a single photon or an ALP decaying promptly to a pair of photons. The particles are emitted $1.5~\mathrm{m}$ from the front face of the calorimeter, approximately the same distance as that between the ATLAS beamline and the front face of the EM calorimeter. To avoid biases from non-uniformities introduced by the accordion geometry, the impact position is randomized to uniformly cover a $2\times 2$ block of middle layer cells. Figure~\ref{fig:caloSim} illustrates an example EM shower from an ALP decaying to two photons, showing the energy deposited across the three longitudinal layers of the EM calorimeter and the first layer of the hadronic calorimeter.

The MC samples are simulated to uniformly cover ALP masses in the range from 10~MeV to 2.5~GeV, as well as ALP and single photon momentum values in the range from 50~GeV to 300~GeV. These values were chosen to provide varying levels of collimation between the photon pair from the ALP decay, which can be characterized as the $\Delta R_{\gamma\gamma}$ angular distance between the two photons. The approximate relationship between this angular distance and the ALP mass and momentum, $m_a$ and $p_{\mathrm{T},a}$ respectively, is given by:

\begin{equation}
    \label{eq:dR}
    \Delta R_{\gamma\gamma} \equiv \sqrt{\left(\Delta \eta_{\gamma\gamma}\right)^2 + \left(\Delta\phi_{\gamma\gamma}\right)^2} \approx \frac{2m_a}{p_{\mathrm{T},a}}.
\end{equation}

\noindent For the mass and momentum ranges described, the samples generated cover $\Delta R_{\gamma\gamma}$ values from approximately $10^{-4}$ to $10^{-1}$. The finest granularity of the calorimeter corresponds to an angular scale of roughly $\Delta R \sim 3\times10^{-3}$, such that much of this regime probes photon-jets with separations smaller than a single calorimeter cell. 
Furthermore, the $\Delta R_{\gamma\gamma}$ values in this sample are up to two orders of magnitude smaller than those considered in similar studies previously~\cite{ai2024detectinghighlycollimatedphotonjets}, representing a significantly more challenging classification dataset. A total of 1 million signal ALP events and 1 million background single photon events are generated, giving 2 million events produced overall. Additional details on the simulated samples are provided in Appendix~\ref{sec:sim_samples}.


%% file: sections/model.tex
\section{Machine Learning Models}
\label{sec:model}

To assess the gain in performance from a cell-level approach and the use of Transformer and MLP Mixer models, a set of benchmark ML models is constructed based on established methods in the literature. 
These models span both high-level approaches, using predominantly SSVs as inputs, and low-level approaches using calorimeter cell information directly. Four benchmark models are considered, namely a BDT, DNN, CNN, and PFN~\cite{Komiske_2019}. While previous literature has leveraged these architectures in varying scenarios, here the models are designed and optimized for the simulation described in Section~\ref{sec:samples} to provide equal context for comparison.

The BDT and DNN in this study were trained using SSV inputs only, as defined in Appendix~\ref{sec:ssvs}. The definition of these variables are chosen from those well-established for describing EM shower development in the ATLAS calorimeters, which are widely used in photon and electron identification, as well as in EM object calibrations~\cite{egamma_energy_calib, ATLAS:2016xha, photonidatlas_2019, electron_id}. The remaining models, including the CNN, PFN, Transformer, and MLP Mixer, operate on calorimeter cell-level inputs. For each event, these inputs consist of the deposited energy in each cell, together with the cell position, given by the $x$ ($\eta$) and $y$ ($\phi$) coordinates, and the calorimeter layer. At most 200 cells per event are used, which is chosen to approximately accommodate the typical cell occupancy of clusters produced by the ATLAS topological clustering algorithm~\cite{atlas_topoclustering} and accounts for $>$99\% of the total energy deposit of the EM shower. No explicit clustering is applied; instead, cells are ordered by energy such that the highest energy cells, which are those most likely to pass clustering thresholds, are retained. This method results in an input representation of shape $(200, 4)$ for each event. The remaining cells are truncated if the total number of cells exceeds 200; if the number of cells is less than 200, the event is padded to the correct shape with a validity mask. For all models, the simulated dataset is divided into training, validation, and test sets using an 80:10:10 split after random shuffling. This split corresponds to approximately 1.6 million events for training and 200,000 events each for validation and testing, with equal fractions of photon-jets and single photons.

\subsection{Benchmark Architectures}
\label{subsec:benchmark_models}

Each benchmark model tested was optimized independently for photon-jet classification with independent hyperparameter scans. Any differences in preprocessing, input representation, and training configuration reflect the need to provide for the distinct requirements of each architecture.

The BDT is implemented using the XGBoost framework~\cite{Chen_2016} with a gradient boosting configuration of 100 trees, a maximum depth of 5, and a learning rate of 0.1. To mitigate overfitting, subsampling is applied at both the event- and feature-level, with fractions of 0.8 used for each. The model is trained using a binary logistic objective function, and evaluated during training using classification error, log-loss, and the root-mean-square error (RMSE) as auxiliary metrics. The number of trees and the maximum depth were optimized via hyperparameter scans, and the model chosen was the most performant.

The DNN is implemented in PyTorch~\cite{pytorch} as a fully-connected feed-forward network, taking the 13 SSVs as inputs. The architecture consists of three hidden layers of 32 nodes each, with ReLU activations applied after each hidden layer, and a sigmoid activation on the single output node to produce a binary classification score. The network is trained for 50 epochs using the Adam~\cite{kingma2017adammethodstochasticoptimization} optimizer with a learning rate of $10^{-3}$ and a batch size of 5000, minimizing a binary cross-entropy loss. Input features are standardized prior to training using a scaler fitted on the training set. The model dimensions and batch size were optimized via hyperparameter scans.

The CNN is implemented in PyTorch using the \texttt{spconv} library~\cite{spconv2022}, which exploits the sparsity of calorimeter data by performing convolutions only over occupied cells, following the submanifold sparse convolution approach~\cite{SubmanifoldSparseConvNet}. Each event is represented on a $144 \times 64$ spatial grid in native detector coordinates, with the first three EM calorimeter layers treated as input channels; the remaining layers are found to contribute no improvement in classification performance and were excluded to reduce training time. The convolutional backbone consists of three sparse convolution blocks with $5\times5$, then $3\times3$ kernels, progressively increasing feature depth from 32 to 128 channels, each followed by batch normalization, ReLU activation, and $2\times2$ max pooling, before the sparse tensor is converted to a dense representation and flattened. The model is trained to perform photon-jet classification using a binary cross-entropy loss, with the model parameters and input representation optimized via hyperparameter scans. The cell $x$ and $y$ positions are provided relative to the EM shower barycenter\footnote{The barycenter is calculated as the energy-weighted position of the shower in the calorimeter~\cite{atlas_topoclustering}. This value is given by $\vec{s} = \sum_{i} E_i\vec{x}_i / \sum_i E_i$, where $E_i$ and $\vec{x}_i$ represent the cell energy and position, respectively, and $i$ runs over the calorimeter cells.}. The model is trained using the Adam optimizer with a learning rate of $10^{-3}$, a batch size of 16, and mixed-precision training, with learning rate reduction on plateau and early stopping based on the validation loss.

The PFN is implemented in Keras~\cite{chollet2015keras} and TensorFlow~\cite{tensorflow}, operating on calorimeter cells as a variable-length, permutation-invariant point cloud. Each cell is described by the same four components $(E, x, y,\mathrm{layer})$ used by the other cell-level models, with zero-padded cells masked and excluded from the computation. 
The PFN can be mathematically summarized as
\begin{equation}
    \mathrm{PFN} = F\left(\sum_{i=1}^{M}\Phi(p_i) \right),
\end{equation}

\noindent where $p_i$ contains the ``particle-level" information about particle $i$, here used to model a calorimeter cell. A shared function $\Phi$ transforms individual particle features into latent representations that are aggregated in a permutation-invariant manner, after which a function $F$ operates on the combined latent representations to perform binary classification. Both $\Phi$ and $F$ are realized as networks of 6 fully-connected layers of width 128 with ReLU activations. The model is trained using the Adam optimizer with an initial learning rate of $2\times10^{-4}$, a batch size of 256, and categorical cross-entropy loss, with learning rate reduction on plateau and early stopping applied during training. The cell $x$ and $y$ positional information is provided to the model as values relative to the EM shower barycenter. Input cell energies are standard scaled, while the $x$ and $y$ positional information are scaled to $[-1,1]$ using a \texttt{MinMaxScaler}, both fitted on the training set. The cell layer information is not scaled, but instead treated as integer encoded. A hyperparameter scan was performed to optimize the dimensions of the network, while keeping the model complexity similar to architectures previously implemented in the literature~\cite{ai2024detectinghighlycollimatedphotonjets}.

\subsection{Transformer}
\label{subsec:Transformer_model}

The Transformer is implemented using the SALT framework~\cite{Barr:2025djz, atlas_salt_docs_2025} originally designed for ATLAS flavor tagging~\cite{ATL-PHYS-PUB-2022-027, atlascollaboration2025transformingjetflavourtagging}, which provides a flexible interface to train multi-modal, multi-task, Transformer-based architectures. The Transformer leverages a self-attention mechanism~\cite{vaswani2023attentionneed}, in which each cell attends to all other cells in the event via learned query $(Q)$, key $(K)$, and value $(V)$ projections of the inputs, as described by the similarity measure,

\begin{equation}
    \mathrm{Attention}(Q, K, V) = \mathrm{Softmax}\left(\frac{QK^\top}{\sqrt{d_k}}\right)V,
\end{equation}

\noindent where $d_k$ is the key dimension, and the scaling $\sqrt{d_k}$ is included to prevent the dot products from growing large with increasing $d_k$. Several of these attention layers can be combined for multi-headed self-attention (MHSA), where each of the $Q,K,$ and $V$ is learned in parallel for $H$ attention heads, and then aggregated via concatenation. MHSA allows the Transformer to capture pairwise correlations between cells in a robust way, with each attention head free to attend to different features of the shower. The attention operation naturally enforces permutation invariance, since the $QK^\top$ dot product is calculated pairwise between all cells symmetrically. Variable-length inputs are similarly accommodated, as the attention matrix adapts to however many cells are present in a given event. These properties provide key conceptual advantages for cell-level modeling with a Transformer compared to the fixed spatial structure imposed by the CNN, or the global pooling operations employed by the PFN.

\begin{figure}[tbh]
\centering
\includegraphics[width=1\textwidth]{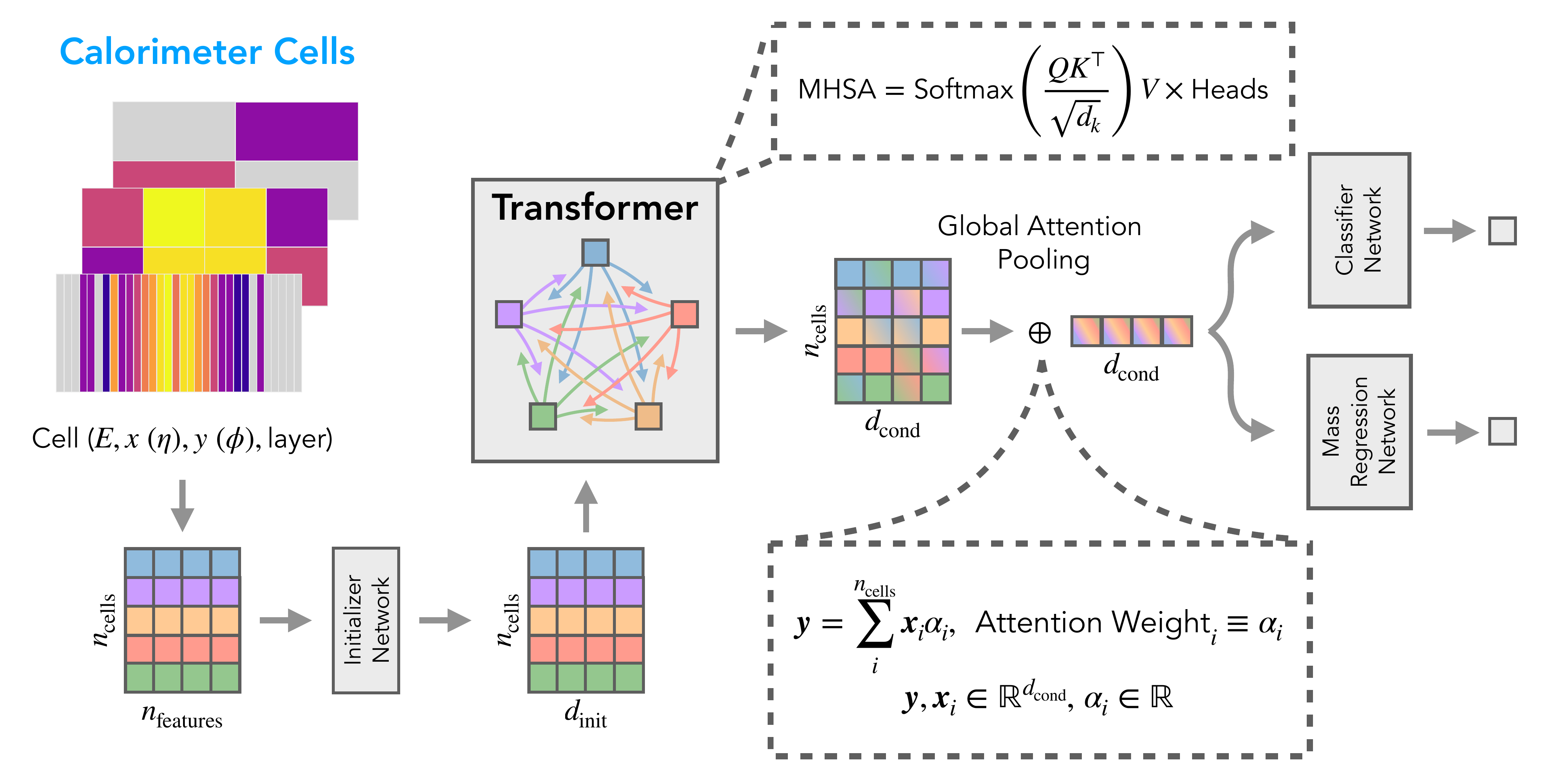}
\caption{\label{fig:arch_trans} Architecture of the Transformer model used in this study. In the diagram, $n_{\mathrm{features}} =4$ and $n_{\mathrm{cells}}=200$, matching the input dimension of the cell dataset. Both the initialization network dimension, $d_{\mathrm{init}}$, and the representation dimension after the Transformer encoder, $d_{\mathrm{cond}}$, are 128. The global attention pooling is calculated as a weighted sum over the cell representations, using attention weights $\alpha_i$ per cell. The output of the pooling is used as input for both the classifier and mass regression task heads.}
\end{figure}

Figure~\ref{fig:arch_trans} shows a schematic diagram of the Transformer architecture used in this study. Each cell is first projected into a 128-dimensional embedding space by an initialization network consisting of a single hidden layer of width 256 with SiLU activations~\cite{elfwing2017sigmoidweightedlinearunitsneural}. These per-cell embeddings are then processed by a Transformer encoder consisting of 4 layers and $H=8$ attention heads. The Transformer has an embedding size of 128 and a feed-forward dimension of 256, and uses pre-LayerNorm~\cite{xiong2020layernormalizationTransformerarchitecture} for training stability. Input features are standard scaled on-the-fly using a precomputed normalization dictionary fitted on the training set. Following the encoder, the output cell representations are projected to a dimension of 128, and a global attention pooling layer~\cite{li2017gatedgraphsequenceneural} aggregates the per-cell embeddings into a single event-level representation by computing a weighted sum. The attention weights $\alpha_i$ used are themselves learned, allowing the model to focus on the most discriminating cells when forming the global representation.

As in the other cell-level methods, each cell is described by $(E, x, y, \mathrm{layer})$. Similarly to the PFN, the cell $x$ and $y$ positions are provided to the Transformer relative to the EM shower barycenter, and the zero-padded cells are masked and excluded from the computation. 
The pooled representation is then passed to two independent task heads, each consisting of fully-connected layers of dimensions 128, 64, and 32, with SiLU activations and a dropout rate of 0.065. 
The first performs binary classification of signal versus background using a cross-entropy loss. 
The second is a regression of the diphoton invariant mass directly from the calorimeter cell information, similar to previous approaches presented in the literature~\cite{Tumasyan_2023, PhysRevLett.134.041801, PhysRevLett.131.101801, cmscollaboration2026searchexotichiggsboson}. Unlike these previous methods, this Transformer combines both tasks in one training and is intended to provide an end-to-end approach for reconstructing the photon-jet object, where the mass regression feeds back into the classification and vice versa.
This second task head uses a LogCosh loss~\cite{saleh2024statisticalpropertieslogcoshloss}, chosen due to its robustness to large residuals, given that the regression target spans three orders of magnitude across the ALP mass spectrum. Single photons enter the mass regression with a target mass value of zero. The two losses are combined as $\mathcal{L} = \mathcal{L}_{\mathrm{classifier}} + 4\mathcal{L}_{\mathrm{regression}}$, where the relative weight of 4 is chosen empirically such that both losses contribute approximately equally during training. The model is trained using the AdamW~\cite{loshchilov2019decoupledweightdecayregularization} optimizer, warming up to a maximum learning rate of $5\times10^{-4}$ before decaying to $1\times10^{-5}$, with a weight decay of $1\times10^{-5}$ and a batch size of 128, for up to 50 epochs. A hyperparameter scan was performed to optimize the architecture, varying the embedding dimension, number of attention heads, number of encoder layers, task head dimensions, and dropout rates, as well as the relative weighting of the classification and regression losses.






\subsection{MLP Mixer}
\label{subsec:mlpmixer}

The MLP Mixer~\cite{tolstikhin2021mlpmixerallmlparchitecturevision} is implemented in Keras as a lightweight alternative to attention-based architectures, relying instead on alternating token-mixing and channel-mixing MLPs. Unlike convolutional or attention-based models, MLP Mixers rely primarily on fully-connected layers applied independently across spatial and channel dimensions. This approach results in predictable data flow, lower control overhead, and a high degree of parallelism, which maps efficiently to FPGA architectures. As a result, MLP Mixer models have begun to be explored in high-energy physics for fast trigger tasks that also require the inference capability of more complex ML architectures~\cite{Sun:2930164}.

\begin{figure}[tbh]
\centering
\includegraphics[width=0.95\textwidth]{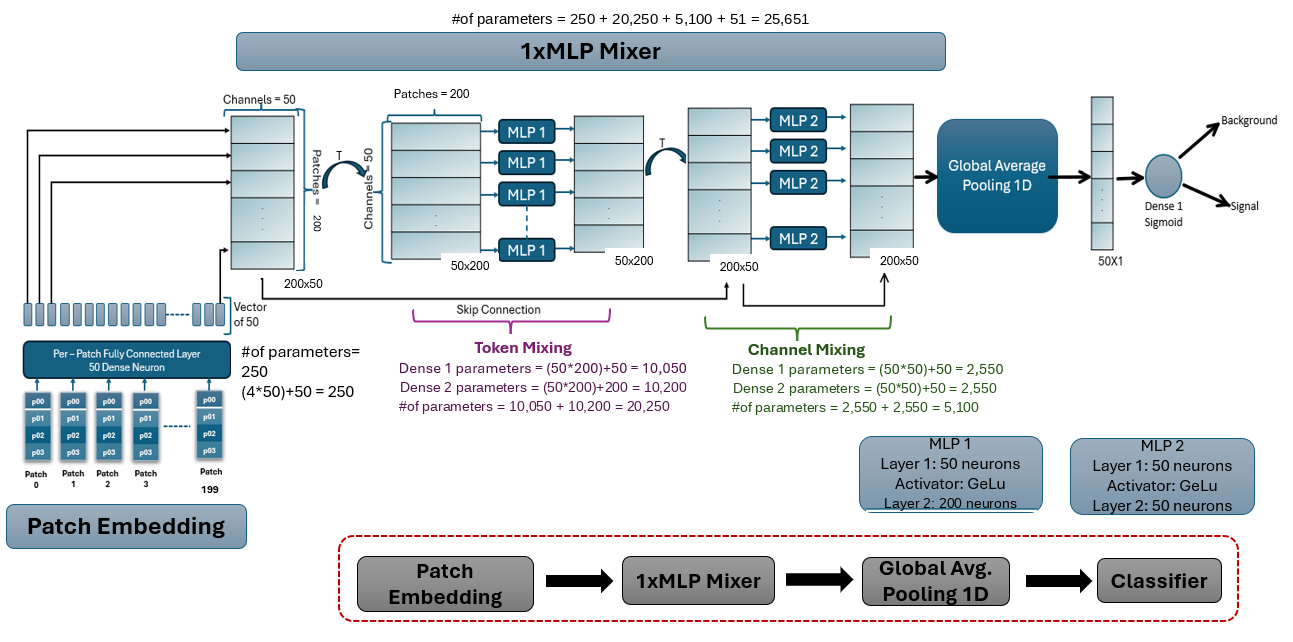}
\caption{\label{fig:arch_mlpm} Architecture of the MLP Mixer model used in this study. The input signal is partitioned into 200 patches, each embedded with a per-patch fully connected layer into a 50-dimensional feature vector (yielding a 50×200 patch-token matrix). A single MLP Mixer block then alternates token mixing (MLP applied across patches for each channel) and channel mixing (MLP applied across channels for each patch). The resulting sequence is aggregated with 1D global average pooling and passed to a sigmoid classifier to predict signal vs. background. Parameter counts for each component are annotated in the diagram.} 
\end{figure}


As in the other cell-level models, events are represented as a set of 200 calorimeter cells, each described by the four features $(E, x, y, \mathrm{layer})$. The MLP Mixer is designed to operate with minimal preprocessing, with raw cell features passed directly to the model without any scaling or normalization applied to the inputs. A diagram of the MLP Mixer architecture used in this study is shown in Figure~\ref{fig:arch_mlpm}. The input cells are partitioned into non-overlapping patches of size 1, such that each cell is treated as an independent token. Each token is projected to an embedding dimension of 50 and passed through a single mixer layer composed of alternating token-mixing and channel-mixing MLPs, where each dense layer in the token- and channel-mixing has a width of 50. The token-mixing MLP operates across the spatial dimension, allowing information to be exchanged between cells, while the channel-mixing MLP operates independently on each cell's feature dimension. The resulting embeddings are aggregated via global average pooling and passed to a single output neuron with sigmoid activation for binary classification. The model is trained using the Adam optimizer with binary cross-entropy loss, a batch size of 32, for up to 200 epochs with early stopping based on validation AUC with a patience of 10 epochs.

%% file: sections/results.tex
\section{Results}
\label{sec:results}

This section summarizes the results achieved by the architectures tested, with the performance for each model assessed by its ability to discriminate photon-jets from the single photon background. First, we focus on the Transformer and MLP Mixer classification performances, with a description of the Transformer mass regression performance included. Finally, a brief comparison between the classification power of these two architectures and the benchmark models is provided at the end of the section.


\subsection{Transformer}
\label{subsec:Transformer}

\begin{figure}[tbh]
    \centering
    \includegraphics[width=0.6\linewidth]{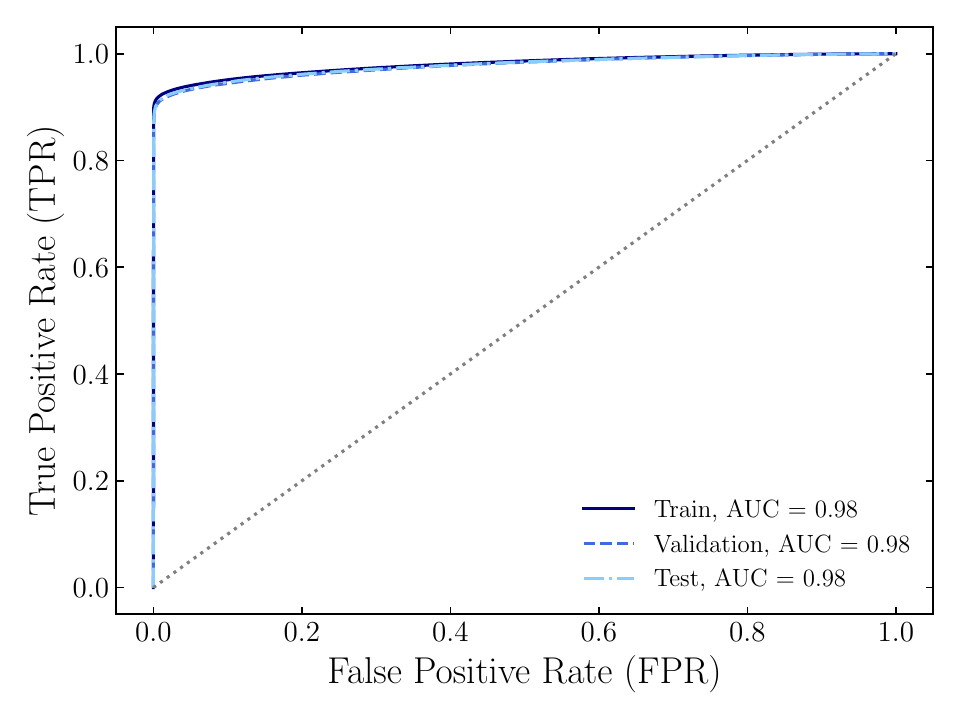}
    \caption{Transformer performance for signal inclusive in ALP mass, showing ROC curves for the train, validation, and test sets. Negligible difference in the AUC is seen between each ROC curve, indicating no signs of overtraining.}
    \label{fig:Transformer_roc}
\end{figure}

The performance of the Transformer is quantified by its receiver operating characteristic (ROC) area-under-curve (AUC), computed inclusive of the ALP masses, as well as in ALP mass bins to assess the performance for varying levels of photon pair collimation. 
Figure~\ref{fig:Transformer_roc} shows the ROC curve for the training, validation, and test sets, plotted inclusively for all signal masses simulated. The Transformer delivers an overall AUC of 0.98 on the test set, demonstrating excellent classification performance. No significant difference in the AUC between the training, validation, and test sets is observed, showing no signs of overtraining of the model. 

\begin{figure}[tbh]
\centering
\includegraphics[width=0.49\textwidth]{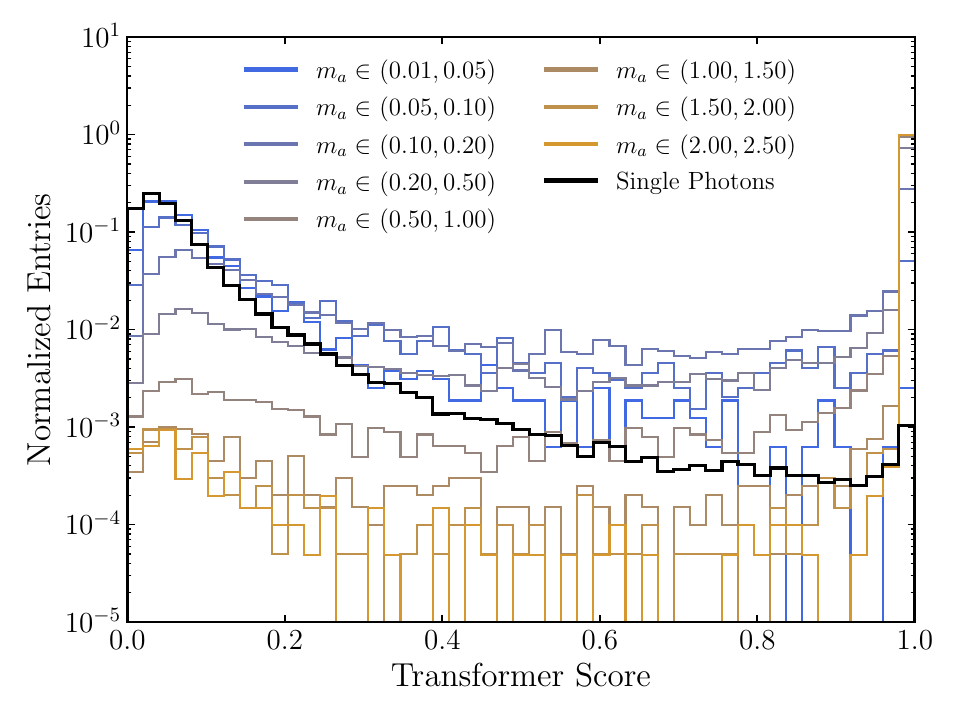}
\includegraphics[width=0.49\textwidth]{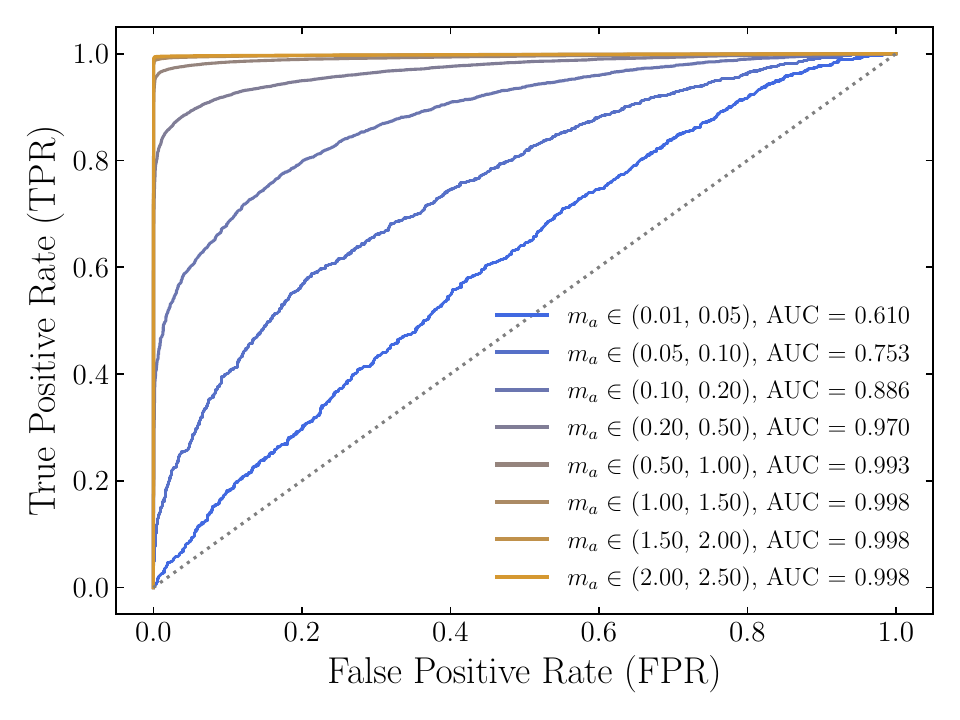}
\caption{\label{fig:Transformer_results}
Transformer performance showing the output classifier score distribution for single photons and signal (left) and the corresponding ROC curves for the test set (right). The signal is broken down into various bins of the ALP mass, shown in units of GeV.}
\end{figure}

\begin{figure}[tbh]
\centering
\includegraphics[width=0.6\textwidth]{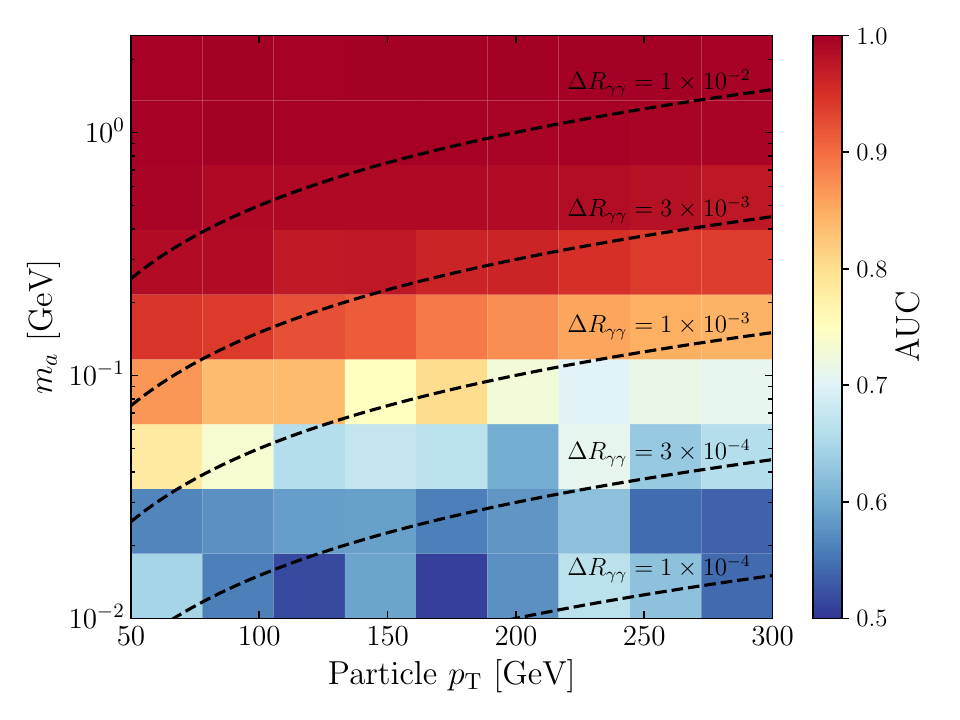}
\caption{\label{fig:Transformer_boost} Transformer AUC as a function of the ALP mass and momentum. Several bands of $\Delta R_{\gamma\gamma}$ are shown, corresponding to different levels of collimation of the photon pair producing the EM shower. A clear degradation in the AUC can be seen for $\Delta R_{\gamma\gamma}$ values below $\mathcal{O}(10^{-3})$, corresponding to separations smaller than the finest calorimeter cell granularity.}
\end{figure}

Figure~\ref{fig:Transformer_results} shows the Transformer output score distributions, with scores ranging from 0 to 1, on the test set for both the ALP signal and single photon background, with the signals separated into eight $m_a$ windows. The ROC curves and AUC values, split into the same eight $m_a$ windows, are also shown. The classification performance varies considerably as a function of ALP mass, expected since lower ALP masses lead to more boosted diphoton decays, which are topologically more similar to single photon showers. For the lowest ALP mass windows of \mbox{10--50~MeV} and 50--100~MeV, the Transformer achieves AUCs of 0.61 and 0.75, respectively. The model retains modest discriminating power in the $\mathcal{O}(10)$ MeV ALP regime, which, due to the difficulty of the photon-jet signature, has not been probed in previous ATLAS or CMS analyses~\cite{PhysRevLett.131.101801, PhysRevLett.134.041801, Tumasyan_2023, PhysRevD.99.012008, haa4y_ATLAS}. Figure~\ref{fig:Transformer_boost} shows the AUC in bins of the generated ALP mass and momentum, with overlaid $\Delta R_{\gamma\gamma}$ bands illustrating the dependence of the classification performance on the degree of collimation. In particular, the lowest mass windows correspond to $\Delta R_{\gamma\gamma}$ separations of $\mathcal{O}(10^{-3})$ and below, smaller than the finest granularity scale of the calorimeter discussed previously. This is an exceedingly challenging regime for a cell-based classifier. Additional studies on the performance of the Transformer as a function of the ALP decay energy asymmetry are included in Appendix~\ref{sec:asymmetry}. 


\begin{figure}[tbh]
\centering
\includegraphics[width=0.6\textwidth]{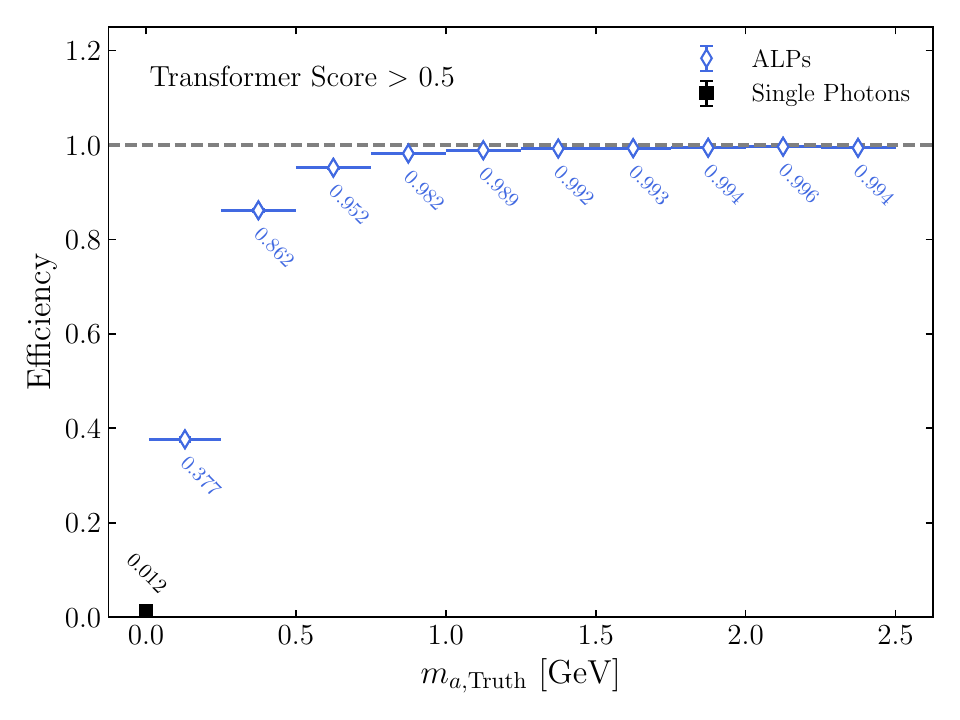}
\caption{\label{fig:transf_sel_eff} Selection efficiency for ALPs and single photons with a Transformer score working point of 0.5, used to assess the mass reconstruction performance. The efficiency numbers for the single photons and each mass bin are annotated in the figure.}
\end{figure}

\begin{figure}[tbh]
\centering
\includegraphics[width=0.48\textwidth]{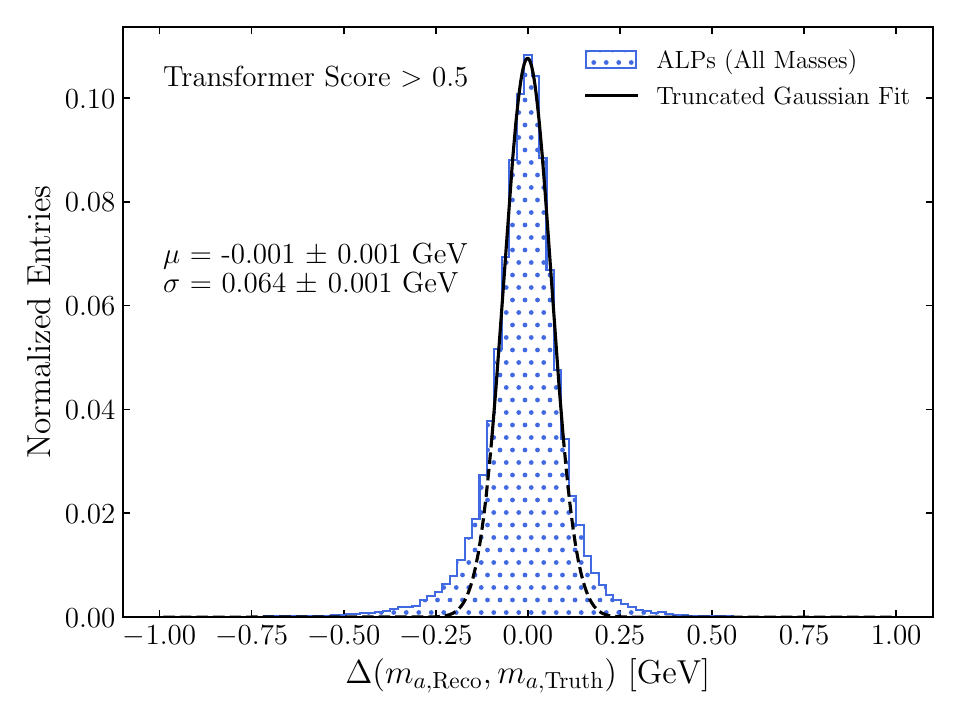}
\includegraphics[width=0.49\textwidth]{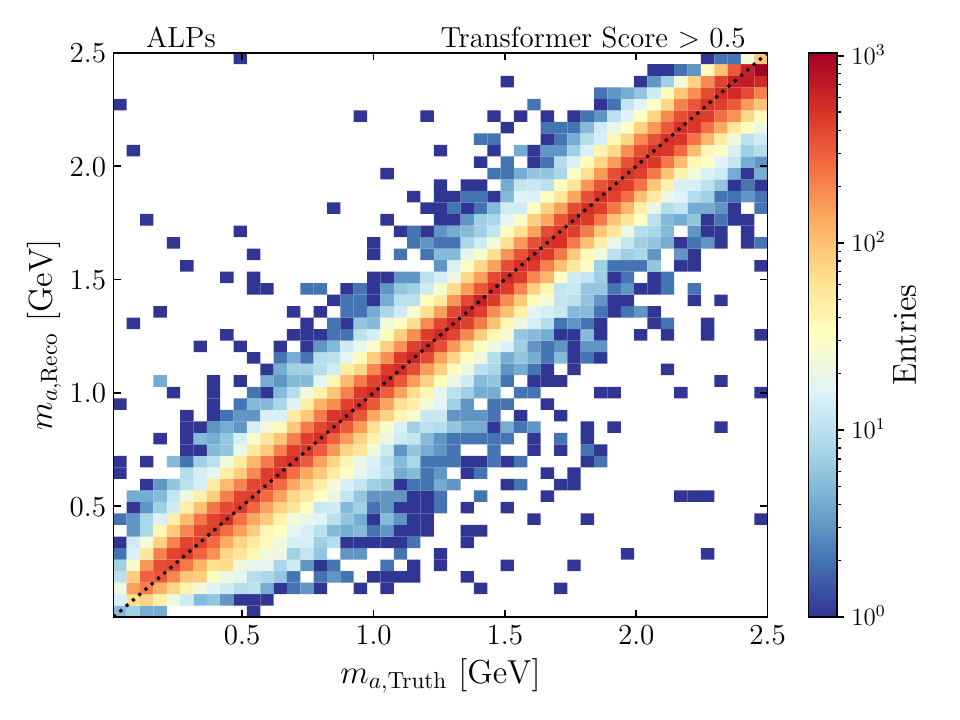}
\caption{\label{fig:transf_alp_mass_reco} Reconstruction error, $\Delta(m_{a,\mathrm{Reco}}, m_{a,\mathrm{Truth}})$, of the diphoton invariant mass provided by the Transformer (left) and linearity between $m_{a,\mathrm{Reco}}$ and $m_{a,\mathrm{Truth}}$ (right). The reconstruction error is fitted to a truncated Gaussian, where the standard deviation is used to determine a mass resolution of $\sigma=64\pm1$~MeV.}
\end{figure}

To assess the mass reconstruction performance, a quality selection cut is employed to prevent any biases on the performance metrics due to fake photons misclassified as ALPs and vice versa. The selection working point is chosen to require true ALPs to have a Transformer score greater than 0.5. The efficiency for this selection is shown in Figure~\ref{fig:transf_sel_eff}, and is $>$80\% for most of the $m_a$ values generated, and $>$95\% for $m_a$ values above 500~MeV. Figure~\ref{fig:transf_alp_mass_reco} shows the reconstruction error residuals, defined as $\Delta(m_{a,\mathrm{Reco}}, m_{a,\mathrm{Truth}}) = m_{a,\mathrm{Reco}}-m_{a,\mathrm{Truth}}$, as well as the linearity between the truth and predicted ALP masses. The mean of the residuals is centered around zero, and no strong non-linearities are seen across the entire generated $m_a$ spectrum, showing good overall reconstruction agreement. To estimate the mass resolution $\sigma_m$ provided by the Transformer, the residual distribution inclusively of all ALP masses is fitted to a Gaussian, with the fit truncated to $\pm 1.5\sigma$ around the mean value to capture only the core of the distribution. The resolution is defined as the standard deviation of the fit, and is determined to be $64 \pm 1$~MeV. 

\begin{figure}[tbh]
\centering
\includegraphics[width=0.49\textwidth]{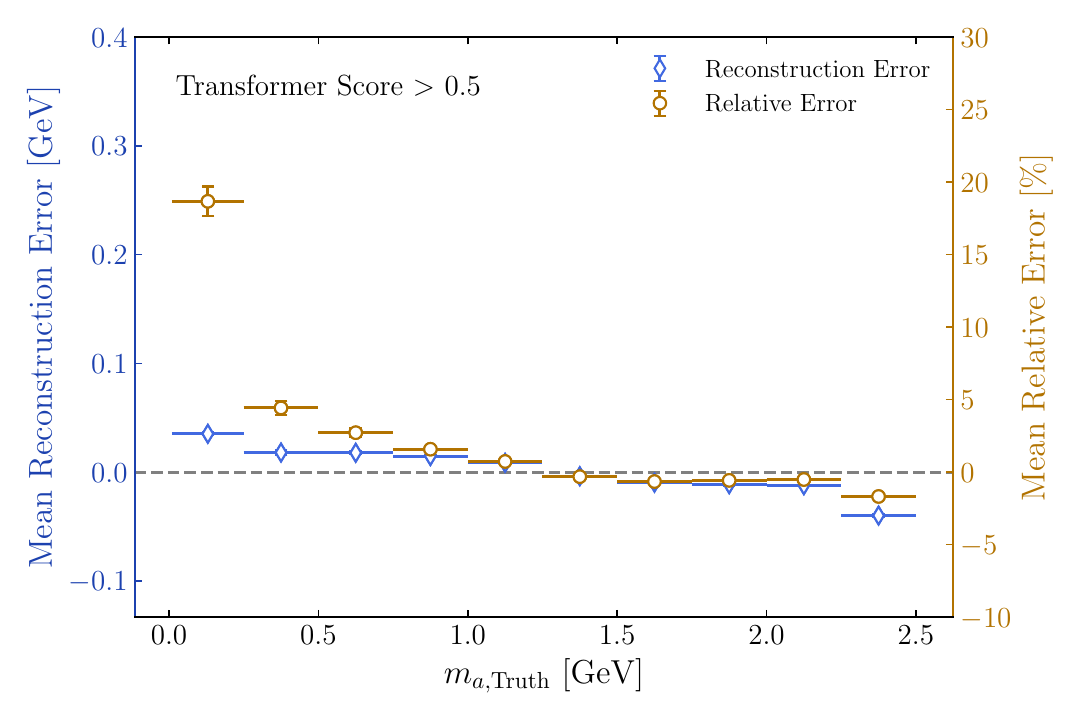}
\includegraphics[width=0.49\textwidth]{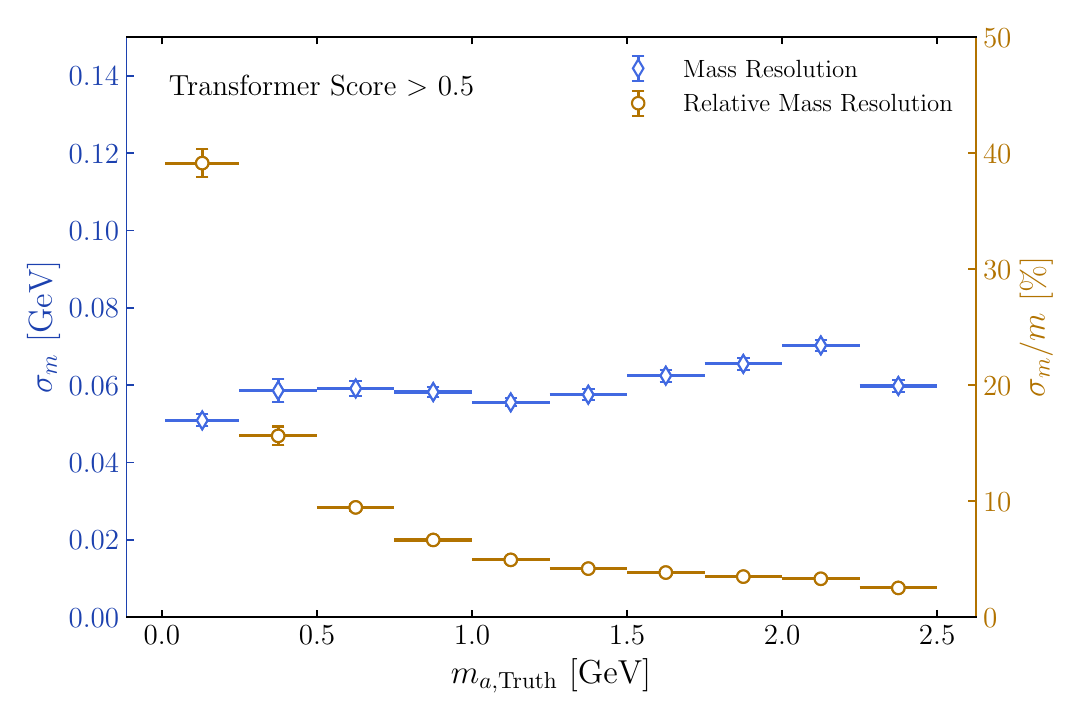}
\caption{\label{fig:transf_alp_mass_reco_by_mass} Mean and relative reconstruction error (left) and absolute and relative mass resolutions (right), in bins of $m_{a, \mathrm{Truth}}$. The relative error is predominantly within 5\%, and the relative resolution is below 10\% for most masses shown.}
\end{figure}

The procedure above is also repeated in bins of $m_a$, where both the mean of the reconstruction and relative\footnote{The relative error here is defined as $\Delta(m_{a,\mathrm{Reco}}, m_{a,\mathrm{Truth}})/m_{a,\mathrm{Truth}}$.} errors are calculated and plotted as a function of $m_a$. The results are shown in Figure~\ref{fig:transf_alp_mass_reco_by_mass}, where the reconstruction error is contained within 50~MeV, and the relative error is predominantly within 5\%. Some reconstruction performance degradation is observed towards the very low and high edges of the ALP masses generated, as the model has no examples to learn from past these boundaries. We found this behavior to be generally independent of the shape of the training prior, where, for example, training on a sample with exponentially more low-mass ALPs showed similar conclusions. This effect, however, can be mitigated in various ways: at the top end, by training over an extended set of masses and then applying the model on a truncated range of that set, such that evaluation is performed away from the edge effects; and at the bottom end, by techniques like domain continuation, as employed by the CMS deep-learning merged photon reconstruction~\cite{Tumasyan_2023}. The effects observed, however, are small, and further optimization is left for future studies. Figure~\ref{fig:transf_alp_mass_reco_by_mass}, in addition, also shows the mass and relative mass resolutions in bins of $m_a$. The resolution is roughly constant as a function of the ALP mass, at $\approx60$~MeV across the entire mass range, and the relative resolution is below 10\% for most of the values shown. 

\begin{figure}[t]
\centering
\includegraphics[width=0.6\textwidth]{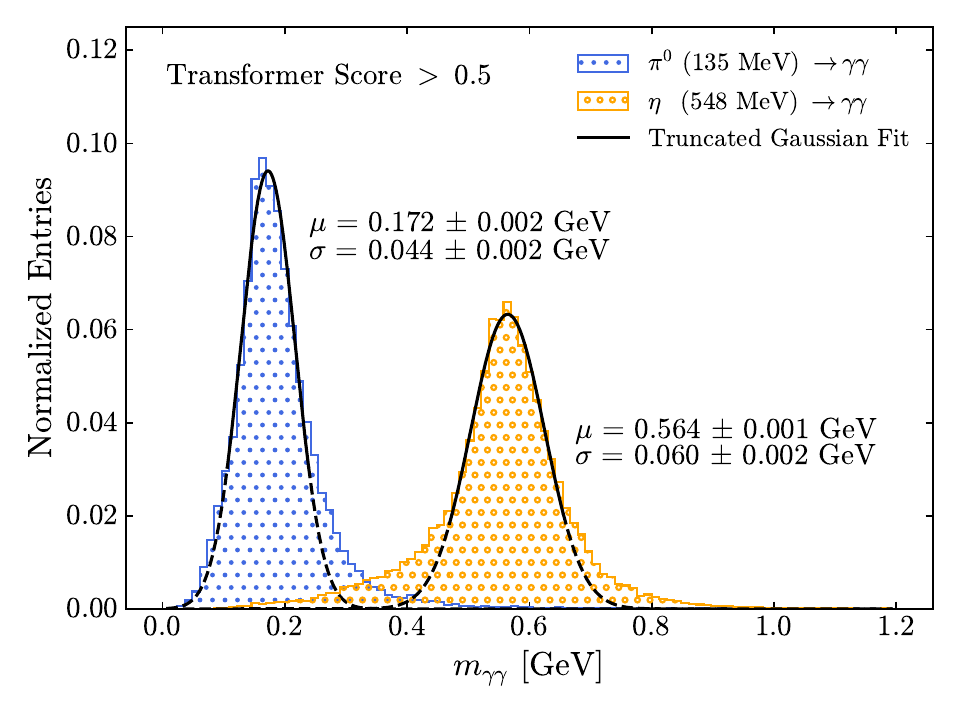}
\caption{\label{fig:transf_pi0_eta_mass_reco} Mass reconstruction of the $\pi^0$ and $\eta$ mesons in the diphoton decay channel.  The reconstructed masses are $m_{\pi^0} = 172\pm2$~MeV and $m_{\eta} = 564\pm1$~MeV, determined from truncated Gaussian fits.}
\end{figure}

As further validation, an additional sample was generated with the particle gun firing both $\pi^0$ and $\eta$ mesons into the calorimeter, with their decays forced to two photons. Figure~\ref{fig:transf_pi0_eta_mass_reco} shows the ability of the trained Transformer to reconstruct the mass distribution of these particles, yielding reconstructed masses of $m_{\pi^0} = 172\pm2$~MeV and \mbox{$m_{\eta} = 564\pm1$~MeV}. An upward bias is observed with respect to the true values of $m_{\pi^0} = 135$~MeV and \mbox{$m_{\eta} = 548$~MeV}, but the deviations are comparable to the mass resolution of the model. Crucially, both mass peaks are well resolved from one another, with a separation of 392~MeV compared to fitted widths of $\sigma_{m(\pi^0)} = 44$~MeV and $\sigma_{m(\eta)} = 60$~MeV. A model of this kind can be used as a means of rejecting $\pi^0$ and $\eta$ fakes in photon identification, or for identifying these particles in other studies by selecting on the mass regressor output. 

\begin{figure}[t]
\centering
\includegraphics[width=0.49\textwidth]{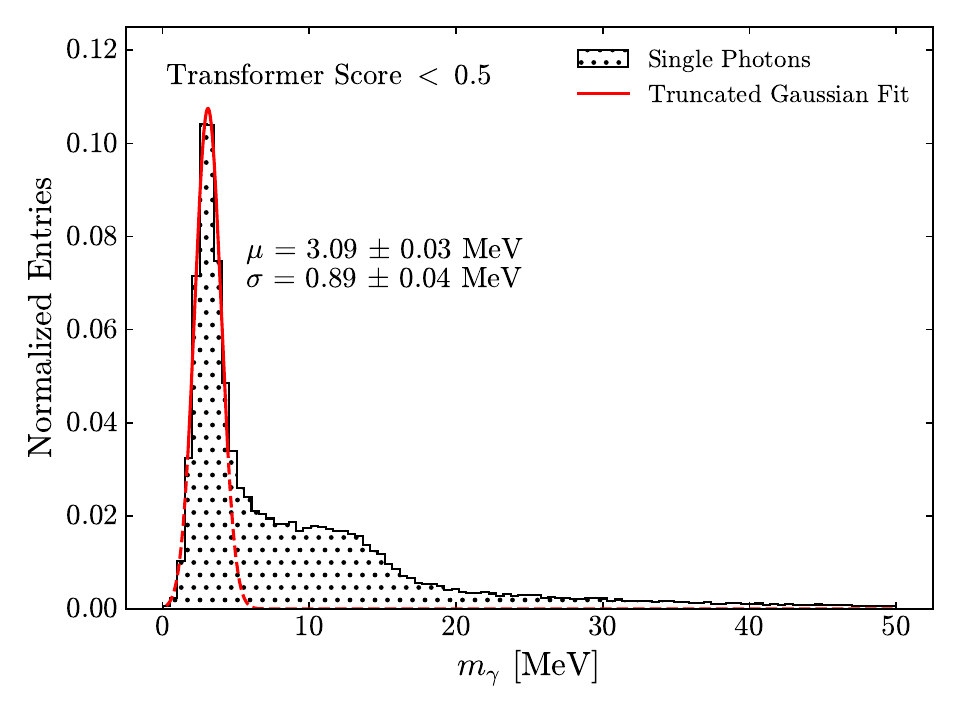}
\includegraphics[width=0.49\textwidth]{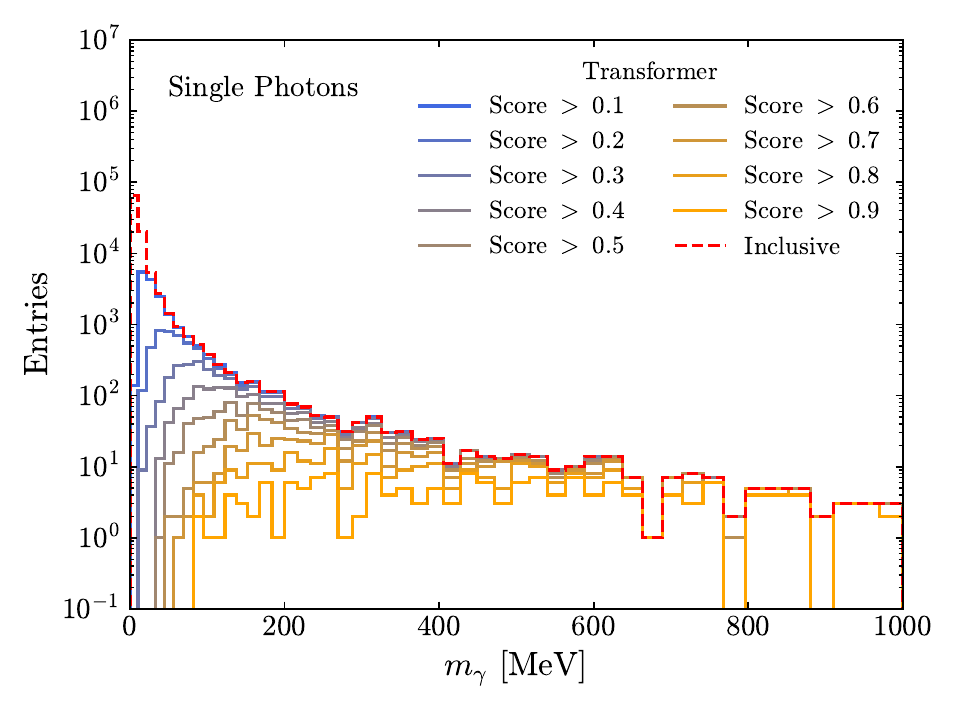}
\caption{\label{fig:transf_photon_mass} Mass reconstruction of single photons using a reversed Transformer score selection (left) and for varying Transformer score thresholds (right). The reversed score selection yields a non-zero reconstructed mass of $m_{\gamma} = 3.09\pm 0.03$~MeV for the massless photon. While some sculpting of the reconstructed $m_{\gamma}$ distribution is observed as the Transformer score threshold varies, the distribution remains smoothly falling in all cases.}
\end{figure}

The reconstructed mass distribution for true single photons passing a reversed Transformer score selection is shown in Figure~\ref{fig:transf_photon_mass}, giving a value of $m_{\gamma} = 3.09\pm 0.03$~MeV. The shoulder on this distribution at around 10--20~MeV can be attributed to a reconstruction error due to photon-jets from very low-mass ALPs closely resembling the shower structure of single photons; this causes an upward bias of the regression value of $m_{\gamma}$ for a fraction of the selected true single photons. Regardless, the overall predicted $m_{\gamma}$ values represent a very small non-zero bias on the reconstructed mass of the massless photon, which is less pronounced than those quoted in similar approaches in the literature~\cite{Tumasyan_2023}. By design, the Transformer score is expected to be correlated with the predicted photon mass, and any decorrelation studies are left for future work. Regardless, Figure~\ref{fig:transf_photon_mass} shows that photons with a varying Transformer score selection provide a smoothly falling distribution, such that this method could be used for a bump-hunt on the Transformer's predicted mass.

\subsection{MLP Mixer}
\label{subsec:mlpmixer}

\begin{figure}[tbh]
\centering
\includegraphics[width=0.49\textwidth]{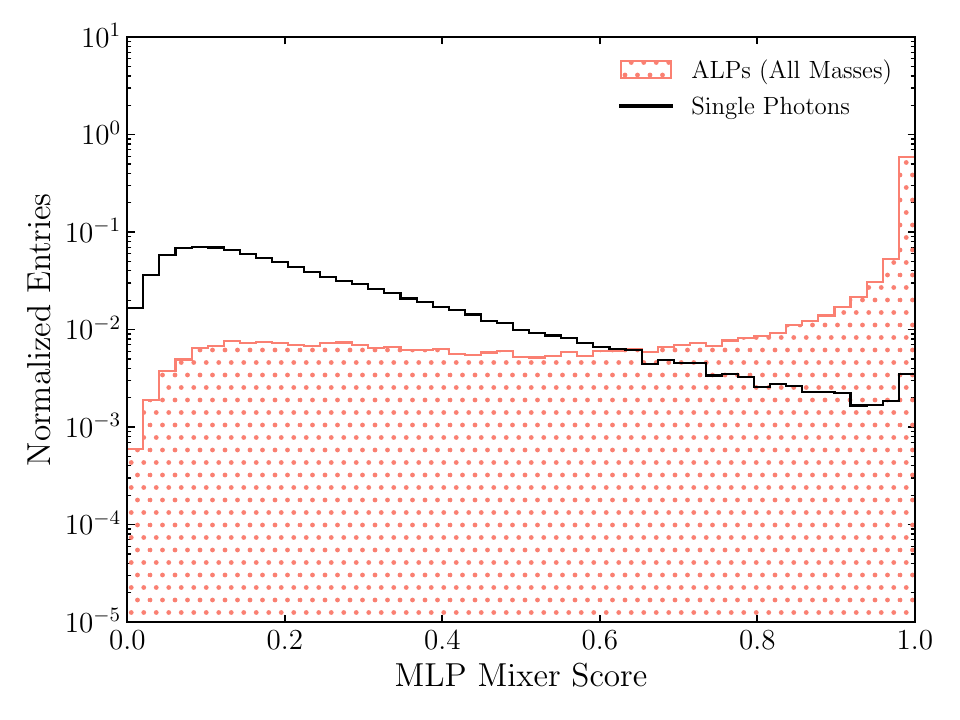}
\includegraphics[width=0.49\textwidth]{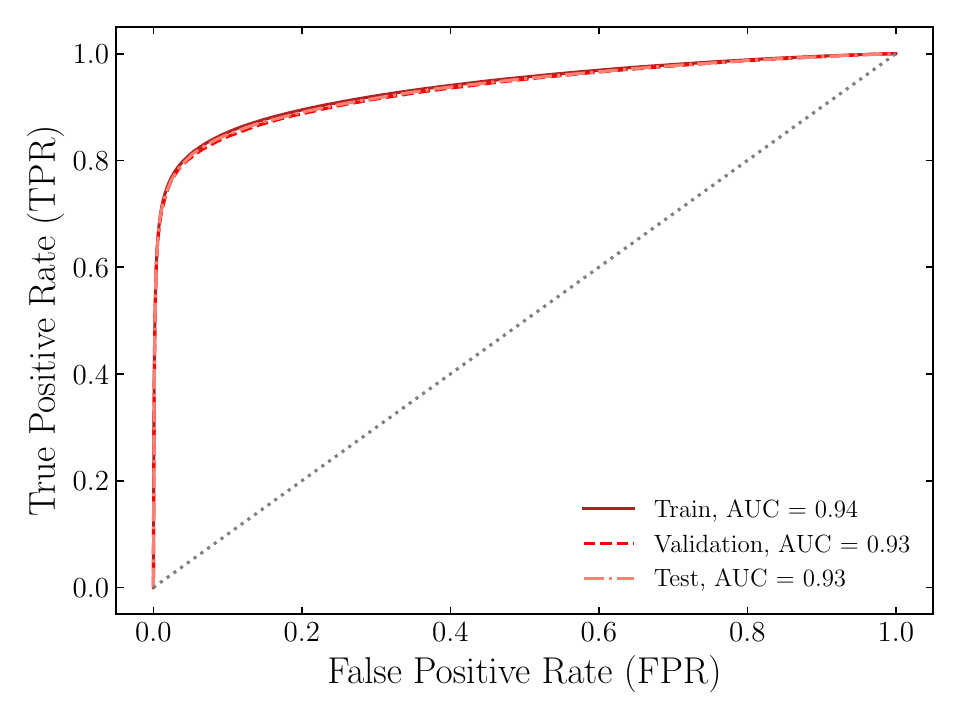}
\caption{\label{fig:mlp_results} Classification performance metrics for the MLP Mixer, specifically the output score distribution for signal and background test events (left) and ROC curve (right), where the signal combines all ALP masses.}
\end{figure}

The signal-background discrimination delivered by the MLP Mixer is shown in Figure~\ref{fig:mlp_results}. 
The output score distributions for the mass-inclusive ALP signal and single photon background test sets are shown, alongside the corresponding ROC curve which yields an AUC of 0.93. 
The reduction in model complexity relative to the Transformer results in a loss of approximately 5\% in AUC. 
This is expected as the MLP Mixer requires a fraction of the computational cost, relying on its alternating token- and channel-mixing structure to capture the relevant correlations in the cell-level inputs without requiring the full pairwise attention computation of the Transformer. 

To fully quantify the computational overhead of each model, the total number of trainable parameters in both the MLP Mixer and Transformer models is tabulated in Table~\ref{tab:mlp_hardware}. 
A comparison of floating-point operations (FLOPs) is also performed, counting specifically the total number of arithmetic operations, including addition, subtraction, multiplication, and division on floating-point numbers. This includes all network layers and attention-value products, but excludes normalization, activation, and softmax operations.
The number of multiply-and-accumulate (MAC) operations, another standard metric for model complexity, is also quoted and can be taken as 50\% of the total number of FLOPs. 
A single inference pass of the MLP Mixer requires $4.08\times 10^6$~FLOPs ($2.04\times10^6$~ MACs), compared to $3.14\times10^8$~FLOPs ($1.57\times10^8$~MACs) for the Transformer, representing a reduction of 99\% in computational cost. The Transformer comprises 634,788 trainable parameters, compared to 25,651 for the MLP Mixer, a reduction of approximately 96\%. The substantial reduction in FLOPs directly translates to lower power consumption and shorter processing times per inference, critical to the deployment of these models in online trigger applications. The decrease in the number of trainable parameters also translates into a decrease in the memory footprint of the model, necessary for deployment on FPGAs. In these ways, the MLP Mixer offers an attractive trade-off between maximum achievable performance, latency and power efficiency requirements, as well as overall memory footprint of potential future cell-level ML trigger algorithms.

\begin{table}[h]
\centering
\begin{tabular}{|c|c|c|}
\hline
 & Transformer & MLP Mixer \\
\hline
Number of  Trainable Parameters & 634,788 & 25,651 \\
Inference FLOPs & 314,167,488 & 4,080,000 \\
Inference MAC Operations & 157,083,744 & 2,040,000 \\
\hline
Total Signal AUC & 0.98 & 0.93 \\
\hline
\end{tabular}
\caption{Comparison of performance (AUC for a combined signal including all simulated mass points), model size (number of trainable parameters), and number of inference FLOPs for the Transformer compared to the MLP Mixer. The MLP Mixer represents an AUC loss of 5\%, but uses only 1\% of the FLOPs of the Transformer model, making it an appealing substitute for a resource-aware inference context. The number of inference MAC operations can be taken to be 50\% of the number of FLOPs.}
\label{tab:mlp_hardware}
\end{table}

\subsection{Comparison with Benchmark Architectures}
\label{subsec:other}

To place the Transformer and MLP Mixer results in broader context, a full comparison of the classification performance across all architectures tested is presented in Table~\ref{tab:comp}, summarizing the mass-inclusive signal AUC for the Transformer and MLP Mixer as well as the benchmark models described in Section~\ref{subsec:benchmark_models}. The numbers presented are averaged over 25 independent trainings with random seeds, and the quoted uncertainties represent the standard deviation across these trainings. The Transformer achieves the highest AUC out of all models tested at 0.9785 ± 0.0003. Particularly relevant to note is that all cell-based architectures outperform the SSV-only models, demonstrating that while SSVs provide meaningful discriminating power, they inherently discard information that is recoverable from the raw calorimeter cell inputs. The consistent improvement seen across all cell-level architectures confirms that moving to low-level, cell representations yields a substantial and general gain in classification performance, irrespective of the specific architecture employed.

\begin{figure}[h]
    \centering
    \includegraphics[width=0.49\linewidth]{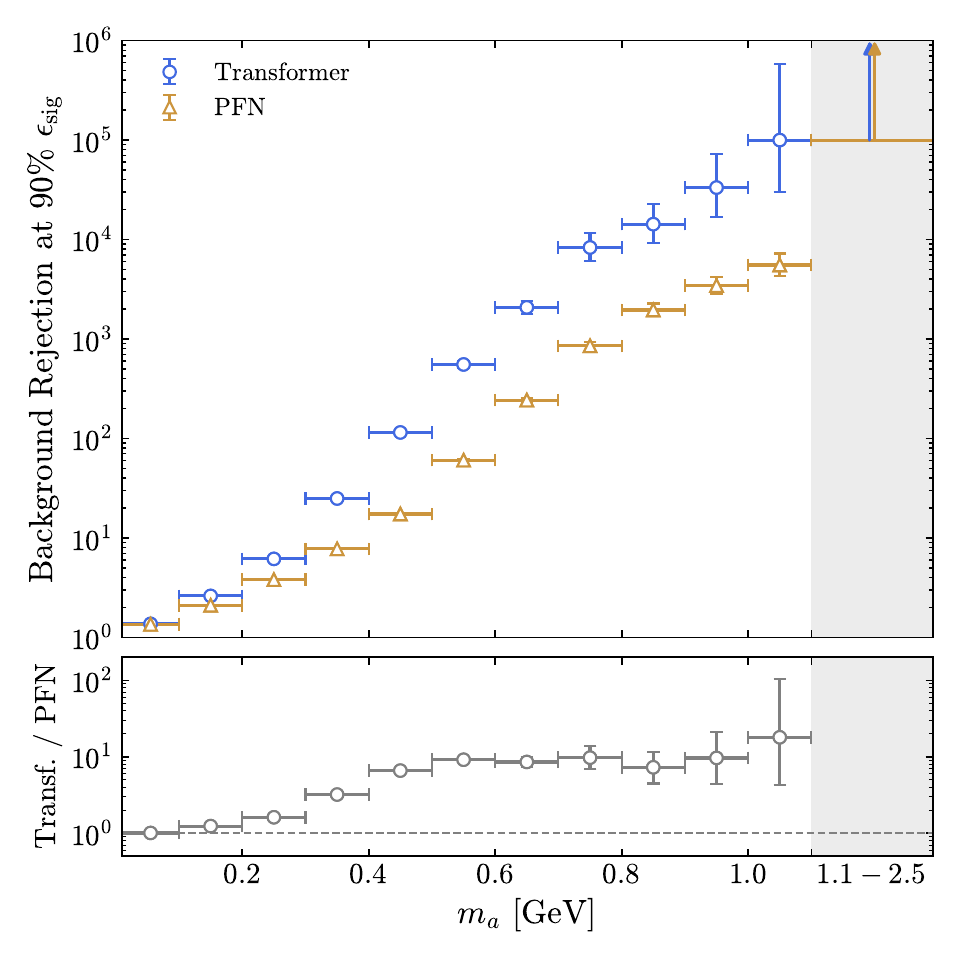}
    \includegraphics[width=0.49\linewidth]{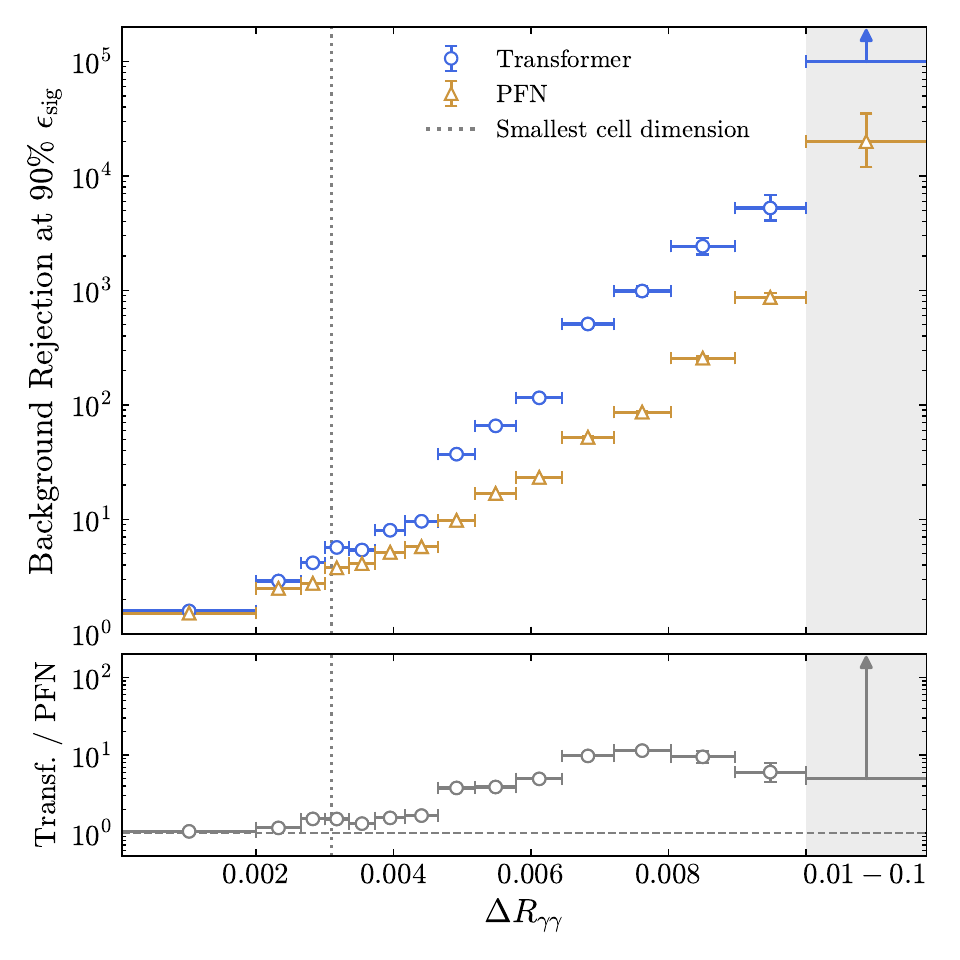}
    \caption{Background rejection at 90\% signal efficiency for the Transformer and PFN as a function of $m_a$ (left) and $\Delta R_{\gamma\gamma}$ (right). The signal efficiency is computed in each $m_a$ and $\Delta R_{\gamma\gamma}$ bin. The arrows pointing to the top of the figure correspond to bins where no background photon survives. The bin value for these is taken as the background rejection for a single photon passing the selection. The shaded region is an overflow bin where either model saturates the background rejection. A dotted line in the $\Delta R_{\gamma\gamma}$ plot shows the finest cell dimension, corresponding to $\Delta\eta = 0.0031$ in the first layer strips. The rejection in the lowest $m_a\in(0.01, 0.1)$~GeV bin is 1.37 and 1.36 for the Transformer and PFN, respectively.}
    \label{fig:bkg_rejection}
\end{figure}

While differences in AUC between the top-performing cell-level models are modest, more pronounced separation emerges when looking at metrics particularly relevant for physics analyses, such as the background rejection at different signal efficiency working points. These numbers are included in Table~\ref{tab:comp}. At 90\% signal efficiency the Transformer achieves a background rejection almost one order of magnitude higher than the PFN, the second-best performing model, and roughly two orders of magnitude higher than the SSV approaches. This is also emphasized in Figure~\ref{fig:bkg_rejection}, where both the Transformer and PFN's background rejection at 90\% signal efficiency are shown as a function of $m_a$ and $\Delta R_{\gamma\gamma}$. Though a comparable improvement is seen in the lowest $m_a$ bin, the Transformer achieves an order of magnitude better background rejection compared to the PFN for most of the ALP mass and diphoton opening angle regime tested. This suggests that, while part of the discriminating power originates from the cell-based modeling, further substantial gains are obtained from the Transformer's self-attention mechanism in extracting a richer shower representation than the remaining benchmark approaches. This enhanced performance is consistent with the Transformer's ability to model pairwise correlations between the cells, beyond what is accessible through global pooling or other cell-level architectures.

To disentangle the Transformer's performance advantage from the multi-task objective, we performed an ablation study with a set of trainings including only the classification task head, with the results also quoted in Table~\ref{tab:comp}. The background rejection values at the various signal efficiency working points are within 30\% of those quoted for the model with the regression head, yet still significantly outperform the PFN and other models tested. This provides further evidence that the performance improvement of the Transformer mostly comes from the richer representations derived from the self-attention mechanism.

Given the 99\% reduction in FLOPs relative to the Transformer, the MLP Mixer provides a compelling architecture where strong performance can be retained with substantially lower computational cost when operating on cell inputs. In terms of model complexity, the CNN, PFN, and Transformer lie in the $\mathcal{O}(10^5$--$10^6)$ parameter range, whereas the SSV-based BDT and DNN contain approximately $\mathcal{O}(10^3)$ parameters. With $\mathcal{O}(10^4)$ trainable parameters, the MLP Mixer occupies an intermediate regime between these extremes, offering a balanced trade-off between model footprint, computational cost, and classification performance for real-time applications while retaining low-level input modeling.


Taken together, the results presented in this study demonstrate that Transformer-based architectures are well-suited to calorimeter cell-level classification. The implementation considered delivers state-of-the-art discrimination across a broad range of ALP masses, including modest performance in the highly collimated, sub-100~MeV regime not previously accessible. The additional mass regression capability from the Transformer, achieving a resolution of \mbox{$\approx 60$~MeV} across the full generated ALP mass spectrum, further underscores the richness of information accessible from the raw calorimeter cells. Furthermore, the MLP Mixer architecture studied provides a complementary alternative for cell-level modeling, with potential for resource-constrained or real-time applications. While the results presented here are based on an idealized simulation of the ATLAS calorimeters, they provide strong motivation for the application of these architectures to real detector data. 




\begin{table*}[hbpt]
\centering
\renewcommand{\arraystretch}{1.15}
\begin{tabular}{|p{35mm}|c|c|c|c|}
\cline{3-5}
\multicolumn{2}{c|}{} & \multicolumn{3}{c|}{Background Rejection ($1/\epsilon_{\mathrm{bkg}}$)} \\
\hline
Model & Overall AUC & 90\% $\epsilon_{\mathrm{sig}}$ & 95\% $\epsilon_{\mathrm{sig}}$ & 99\% $\epsilon_{\mathrm{sig}}$ \\
\hline
\multicolumn{5}{|c|}{SSV-Level} \\
\hline
BDT & 0.9245 ± 0.0001 & 3.22 ± 0.01 & 1.83 ± 0.01 & 1.170 ± 0.004 \\
\hdashline
DNN & $0.923 \pm 0.001$ & $3.1\pm0.1$ & $1.80\pm0.02$ & $1.20\pm0.01$ \\
\hline
\multicolumn{5}{|c|}{Cell-Level} \\
\hline
CNN & $0.939 \pm 0.002$ & $4.7\pm0.3$ & $2.3\pm0.1$ & $1.30\pm0.02$ \\
\hdashline
PFN & 0.9724 ± 0.0005 & $62\pm7$ & $5.4\pm0.2$ & $1.48\pm0.01$ \\
\hdashline
  \textbf{Transformer} \newline \hspace*{1em}(Classification)
  & \textbf{0.9777 ± 0.0006} & {\boldmath$\mathrm{343}\pm \mathrm{64}$} & {\boldmath$\mathrm{8.2}\pm \mathrm{0.5}$} & {\boldmath$\mathrm{1.62}\pm \mathrm{0.02}$} \\
\hdashline
\textbf{Transformer} \newline \hspace*{1em}(Classif. \& regr.)
  & \textbf{0.9785 ± 0.0003} & {\boldmath$474\pm61$} & {\boldmath$9.0\pm0.4$} & {\boldmath$1.64\pm0.01$} \\
\hdashline
\textbf{MLP Mixer} & \textbf{0.939 ± 0.003} & \textbf{5.0 ± 0.4} & \textbf{2.2 ± 0.1} & \textbf{1.19 ± 0.01} \\
\hline
\end{tabular}
\caption{Comparison of signal AUC for BDT, DNN, CNN, PFN, Transformer, and MLP Mixer models. The results for the Transformer are shown for both classification-only and classification-and-regression models. For various signal efficiencies, $\epsilon_{\mathrm{sig}}$, the background rejection is shown as $1/\epsilon_{\mathrm{bkg}}$, where $\epsilon_{\mathrm{bkg}}$ is the background efficiency. For all the models, the quoted AUCs and background rejection values are averaged over 25 independent trainings with different random seeds. The uncertainties correspond to the standard deviation across these trainings. The BDT and DNN models use high-level SSV features, while the CNN, PFN, Transformer, and MLP Mixer use low-level calorimeter cell inputs. 
\label{tab:comp}}
\end{table*}

%% file: sections/conclusions.tex
\section{Conclusions}
\label{sec:conclusions}


This study presents a systematic investigation of ML architectures for the classification of photon-jets from ALP decays against the single photon background, using cell-level representations of an ATLAS-like calorimeter. A Transformer and an MLP Mixer were developed and evaluated as novel approaches to this task, alongside a set of benchmark models spanning both SSV-based and cell-level methods. The Transformer achieves an overall AUC of 0.9785 ± 0.0003, outperforming all benchmark architectures and demonstrating that the self-attention mechanism is particularly well-suited to exploiting the rich spatial and energy correlations encoded in the calorimeter cell data. Modest classification performance is achieved in the highly collimated, sub-100~MeV ALP mass regime, providing some promise for probing this parameter space with these methods in future analyses of particle collider data. In addition to classification, the Transformer demonstrates the ability to perform direct invariant mass regression from the calorimeter cell information, achieving a mass resolution of $64 \pm 1$~MeV.  The MLP Mixer, while sacrificing approximately 5\% in AUC relative to the Transformer, achieves this with only 1\% of the computational cost, offering an attractive alternative for latency- and power-constrained deployment contexts such as real-time trigger systems. These results establish cell-level Transformer-based architectures as a powerful and flexible tool for calorimeter-based particle identification, with applications extending naturally beyond photon-jet classification to a broader class of highly collimated and overlapping shower topologies.

%% file: biblio.bib
@article{PhysRevLett.38.1440,
    author        = {Peccei, R. D. and Quinn, Helen R.},
    title         = {{CP Conservation in the Presence of Pseudoparticles}},
    journal       = {Phys. Rev. Lett.},
    volume        = {38},
    pages         = {1440},
    year          = {1977},
    doi           = {10.1103/PhysRevLett.38.1440}
}

@article{PhysRevD.16.1791,
    author        = {Peccei, R. D. and Quinn, Helen R.},
    title         = {{Constraints Imposed by CP Conservation in the Presence of Instantons}},
    journal       = {Phys. Rev. D},
    volume        = {16},
    pages         = {1791},
    year          = {1977},
    doi           = {10.1103/PhysRevD.16.1791}
}

@article{PhysRevLett.40.223,
    author        = {Weinberg, Steven},
    title         = {{A New Light Boson?}},
    journal       = {Phys. Rev. Lett.},
    volume        = {40},
    pages         = {223},
    year          = {1978},
    doi           = {10.1103/PhysRevLett.40.223}
}

@article{PhysRevLett.40.279,
    author        = {Wilczek, F.},
    title         = {{Problem of Strong $P$ and $T$ Invariance in the Presence of Instantons}},
    journal       = {Phys. Rev. Lett.},
    volume        = {40},
    pages         = {279},
    year          = {1978},
    doi           = {10.1103/PhysRevLett.40.279}
}

@article{Preskill:1982cy,
    author        = {Preskill, John and Wise, Mark B. and Wilczek, Frank},
    title         = {{Cosmology of the Invisible Axion}},
    journal       = {Phys. Lett. B},
    volume        = {120},
    pages         = {127},
    year          = {1983},
    doi           = {10.1016/0370-2693(83)90637-8},
    reportNumber  = {HUTP-82-A048, NSF-ITP-82-103}
}

@article{Abbott:1982af,
    author        = {Abbott, L. F. and Sikivie, P.},
    title         = {{A Cosmological Bound on the Invisible Axion}},
    journal       = {Phys. Lett. B},
    volume        = {120},
    pages         = {133},
    year          = {1983},
    doi           = {10.1016/0370-2693(83)90638-X}
}

@article{Dine:1982ah,
    author        = {Dine, Michael and Fischler, Willy},
    title         = {{The Not So Harmless Axion}},
    journal       = {Phys. Lett. B},
    volume        = {120},
    pages         = {137},
    year          = {1983},
    doi           = {10.1016/0370-2693(83)90639-1},
    reportNumber  = {UPR-0201T}
}

@article{Marsh:2015xka,
    author        = {Marsh, David J. E.},
    title         = {{Axion Cosmology}},
    journal       = {Phys. Rept.},
    volume        = {643},
    pages         = {1},
    year          = {2016},
    doi           = {10.1016/j.physrep.2016.06.005},
    eprint        = {1510.07633},
    archivePrefix = {arXiv},
    primaryClass  = {astro-ph.CO},
    reportNumber  = {KCL-PH-TH-2015-50}
}

@article{Chadha-Day:2021szb,
    author        = {Chadha-Day, Francesca and Ellis, John and Marsh, David J. E.},
    title         = {{Axion dark matter: What is it and why now?}},
    journal       = {Sci. Adv.},
    volume        = {8},
    number        = {8},
    pages         = {eabj3618},
    year          = {2022},
    doi           = {10.1126/sciadv.abj3618},
    eprint        = {2105.01406},
    archivePrefix = {arXiv},
    primaryClass  = {hep-ph},
    reportNumber  = {KCL-PH-TH/2021-20, CERN-TH-2021-045, IPPP/20/91}
}

@article{Mimasu:2014nea,
    author        = {Mimasu, Ken and Sanz, Ver{\'o}nica},
    title         = {{ALPs at Colliders}},
    journal       = {JHEP},
    volume        = {06},
    pages         = {173},
    year          = {2015},
    doi           = {10.1007/JHEP06(2015)173},
    eprint        = {1409.4792},
    archivePrefix = {arXiv},
    primaryClass  = {hep-ph}
}

@article{Jaeckel_2016,
    author        = {Jaeckel, Joerg and Spannowsky, Michael},
    title         = {{Probing MeV to 90 GeV axion-like particles with LEP and LHC}},
    journal       = {Phys. Lett. B},
    volume        = {753},
    pages         = {482},
    year          = {2016},
    doi           = {10.1016/j.physletb.2015.12.037},
    eprint        = {1509.00476},
    archivePrefix = {arXiv},
    primaryClass  = {hep-ph}
}

@article{Bauer_2017,
    author        = {Bauer, Martin and Neubert, Matthias and Thamm, Andrea},
    title         = {{Collider probes of axion-like particles}},
    journal       = {JHEP},
    volume        = {12},
    pages         = {044},
    year          = {2017},
    doi           = {10.1007/JHEP12(2017)044},
    eprint        = {1708.00443},
    archivePrefix = {arXiv},
    primaryClass  = {hep-ph}
}

@article{Agrawal_2021,
    author        = {Agrawal, P. and others},
    title         = {{Feebly-interacting particles: FIPs 2020 workshop report}},
    journal       = {Eur. Phys. J. C},
    volume        = {81},
    pages         = {1015},
    year          = {2021},
    doi           = {10.1140/epjc/s10052-021-09703-7},
    eprint        = {2102.12143},
    archivePrefix = {arXiv},
    primaryClass  = {hep-ph}
}

@article{Yan:2025alk,
    author        = {Yan, Zi-Yao and Feng, Jie},
    title         = {{Axion-like Particle Search with a Light-Shining-Through-Walls Setup at a $\gamma$--$\gamma$ Collider}},
    year          = {2025},
    eprint        = {2512.15192},
    archivePrefix = {arXiv},
    primaryClass  = {hep-ph}
}

@article{Lentz:2026bmp,
    author        = {Lentz, Erik W. and Boutan, Christian R. and Taubman, Matthew S. and Gervais, Kevin L.},
    title         = {{Developing centimeter-scale-cavity arrays for axion dark matter detection in the 100~micro-electron-volt range}},
    journal       = {JINST},
    volume        = {21},
    number        = {04},
    pages         = {P04020},
    year          = {2026},
    doi           = {10.1088/1748-0221/21/04/P04020},
    eprint        = {2601.21074},
    archivePrefix = {arXiv},
    primaryClass  = {hep-ex}
}

@article{Akgumus:2025mrh,
    author        = {Akg{\"u}m{\"u}s, M. A. and Salama, N. and Egge, J. and Garutti, E. and Maroudas, M. and Nguyen, L. H. and Leppla-Weber, D.},
    title         = {{A new limit for axion dark matter with SPACE}},
    journal       = {JCAP},
    volume        = {04},
    pages         = {054},
    year          = {2026},
    doi           = {10.1088/1475-7516/2026/04/054},
    eprint        = {2506.18411},
    archivePrefix = {arXiv},
    primaryClass  = {hep-ex}
}

@article{Knapen_2016,
    author        = {Knapen, Simon and Melia, Tom and Papucci, Michele and Zurek, Kathryn M.},
    title         = {{Rays of light from the LHC}},
    journal       = {Phys. Rev. D},
    volume        = {93},
    pages         = {075020},
    year          = {2016},
    doi           = {10.1103/PhysRevD.93.075020},
    eprint        = {1512.04928},
    archivePrefix = {arXiv},
    primaryClass  = {hep-ph}
}

@article{Chang_2016,
    author        = {Chang, Jung and Cheung, Kingman and Lu, Chih-Ting},
    title         = {{Interpreting the 750 GeV diphoton resonance using photon jets in hidden-valley-like models}},
    journal       = {Phys. Rev. D},
    volume        = {93},
    pages         = {075013},
    year          = {2016},
    doi           = {10.1103/PhysRevD.93.075013},
    eprint        = {1512.06671},
    archivePrefix = {arXiv},
    primaryClass  = {hep-ph}
}

@article{Aparicio_2016,
    author        = {Aparicio, Luis and Azatov, Aleksandr and Hardy, Edward and Romanino, Andrea},
    title         = {{Diphotons from diaxions}},
    journal       = {JHEP},
    volume        = {05},
    pages         = {077},
    year          = {2016},
    doi           = {10.1007/JHEP05(2016)077},
    eprint        = {1602.00949},
    archivePrefix = {arXiv},
    primaryClass  = {hep-ph}
}

@article{ellwanger2016750gevdiphotonsignal,
    author        = {Ellwanger, Ulrich and Hugonie, Cyril},
    title         = {{A 750 GeV Diphoton Signal from a Very Light Pseudoscalar in the NMSSM}},
    journal       = {JHEP},
    volume        = {05},
    pages         = {114},
    year          = {2016},
    doi           = {10.1007/JHEP05(2016)114},
    eprint        = {1602.03344},
    archivePrefix = {arXiv},
    primaryClass  = {hep-ph}
}

@article{Dasgupta_2016,
    author        = {Dasgupta, Basudeb and Kopp, Joachim and Schwaller, Pedro},
    title         = {{Photons, photon jets, and dark photons at 750 GeV and beyond}},
    journal       = {Eur. Phys. J. C},
    volume        = {76},
    pages         = {277},
    year          = {2016},
    doi           = {10.1140/epjc/s10052-016-4127-4},
    eprint        = {1602.04692},
    archivePrefix = {arXiv},
    primaryClass  = {hep-ph}
}

@article{ai2024detectinghighlycollimatedphotonjets,
    author        = {Ai, Xiaocong and Feng, William Y. and Hsu, Shih-Chieh and Li, Ke and Lu, Chih-Ting},
    title         = {{Detecting highly collimated photon-jets from Higgs boson exotic decays with deep learning}},
    year          = {2024},
    eprint        = {2401.15690},
    archivePrefix = {arXiv},
    primaryClass  = {hep-ph}
}

@article{PhysRevD.99.012008,
    author        = {{ATLAS collaboration}},
    title         = {{Search for pairs of highly collimated photon-jets in $pp$ collisions at $\sqrt{s}=13\,\text{TeV}$ with the ATLAS detector}},
    journal       = {Phys. Rev. D},
    volume        = {99},
    pages         = {012008},
    year          = {2019},
    doi           = {10.1103/PhysRevD.99.012008},
    eprint        = {1808.10515},
    archivePrefix = {arXiv},
    primaryClass  = {hep-ex}
}

@article{haa4y_ATLAS,
    author        = {{ATLAS collaboration}},
    title         = {{Search for short- and long-lived axion-like particles in $H\to aa\to 4\gamma$ decays with the ATLAS experiment at the LHC}},
    journal       = {Eur. Phys. J. C},
    volume        = {84},
    pages         = {742},
    year          = {2024},
    doi           = {10.1140/epjc/s10052-024-12979-0},
    eprint        = {2312.03306},
    archivePrefix = {arXiv},
    primaryClass  = {hep-ex}
}

@article{PhysRevLett.131.101801,
    author        = {{CMS collaboration}},
    title         = {{Search for Exotic Higgs Boson Decays $H\to\mathcal{A}\mathcal{A}\to 4\gamma$ with Events Containing Two Merged Diphotons in Proton-Proton Collisions at $\sqrt{s}=13\,\text{TeV}$}},
    journal       = {Phys. Rev. Lett.},
    volume        = {131},
    pages         = {101801},
    year          = {2023},
    doi           = {10.1103/PhysRevLett.131.101801},
    eprint        = {2209.06197},
    archivePrefix = {arXiv},
    primaryClass  = {hep-ex}
}

@article{Tumasyan_2023,
    author        = {{CMS collaboration}},
    title         = {{Reconstruction of decays to merged photons using end-to-end deep learning with domain continuation in the CMS detector}},
    journal       = {Phys. Rev. D},
    volume        = {108},
    pages         = {052002},
    year          = {2023},
    doi           = {10.1103/PhysRevD.108.052002},
    eprint        = {2204.12313},
    archivePrefix = {arXiv},
    primaryClass  = {hep-ex}
}

@article{PhysRevLett.134.041801,
    author        = {{CMS collaboration}},
    title         = {{Search for New Resonances Decaying to Pairs of Merged Diphotons in Proton-Proton Collisions at $\sqrt{s}=13\,\text{TeV}$}},
    journal       = {Phys. Rev. Lett.},
    volume        = {134},
    pages         = {041801},
    year          = {2025},
    doi           = {10.1103/PhysRevLett.134.041801},
    eprint        = {2405.00834},
    archivePrefix = {arXiv},
    primaryClass  = {hep-ex}
}

@article{cmscollaboration2026searchexotichiggsboson,
    author        = {{CMS collaboration}},
    title         = {{Search for exotic Higgs boson decays $H\to\mathcal{AA}$ with $\mathcal{A}\to\gamma\gamma$ in events with a semi-merged topology in proton-proton collisions at $\sqrt{s}=13\,\text{TeV}$}},
    year          = {2026},
    eprint        = {2601.00183},
    archivePrefix = {arXiv},
    primaryClass  = {hep-ex}
}

@article{Evans:2008zzb,
    author        = {Evans, Lyndon and Bryant, Philip},
    title         = {{LHC Machine}},
    journal       = {JINST},
    volume        = {3},
    pages         = {S08001},
    year          = {2008},
    doi           = {10.1088/1748-0221/3/08/S08001}
}

@article{atlasdetectorpaper,
    author        = {{ATLAS collaboration}},
    title         = {{The ATLAS Experiment at the CERN Large Hadron Collider}},
    journal       = {JINST},
    volume        = {3},
    pages         = {S08003},
    year          = {2008},
    doi           = {10.1088/1748-0221/3/08/S08003}
}

@techreport{CERN-LHCC-96-041,
    author        = {{ATLAS collaboration}},
    title         = {{ATLAS Liquid Argon Calorimeter: Technical Design Report}},
    institution   = {CERN},
    address       = {Geneva},
    number        = {CERN-LHCC-96-041},
    year          = {1996},
    doi           = {10.17181/CERN.FWRW.FOOQ},
    url           = {https://cds.cern.ch/record/331061}
}

@techreport{CERN-LHCC-2017-020,
    author        = {{ATLAS collaboration}},
    title         = {{Technical Design Report for the Phase-II Upgrade of the ATLAS TDAQ System}},
    institution   = {CERN},
    address       = {Geneva},
    number        = {CERN-LHCC-2017-020, ATLAS-TDR-029},
    year          = {2017},
    doi           = {10.17181/CERN.2LBB.4IAL},
    url           = {https://cds.cern.ch/record/2285584}
}

@article{Agostinelli:2002hh,
    author        = {Agostinelli, S. and others},
    title         = {{\textsc{Geant4} -- a simulation toolkit}},
    journal       = {Nucl. Instrum. Meth. A},
    volume        = {506},
    pages         = {250},
    year          = {2003},
    doi           = {10.1016/S0168-9002(03)01368-8},
    reportNumber  = {SLAC-PUB-9350, FERMILAB-PUB-03-339}
}

@misc{simplifiedatlassim,
    author        = {Pitt, Michael and Matos, Gabriel},
    title         = {{gabrielpmatos/g4-atlas-calorimeter: Datasets for ML Trainings}},
    howpublished  = {Zenodo},
    year          = {2026},
    doi           = {10.5281/zenodo.18825682}
}

@article{atlas_topoclustering,
    author        = {{ATLAS collaboration}},
    title         = {{Topological cell clustering in the ATLAS calorimeters and its performance in LHC Run 1}},
    journal       = {Eur. Phys. J. C},
    volume        = {77},
    pages         = {490},
    year          = {2017},
    doi           = {10.1140/epjc/s10052-017-5004-5},
    eprint        = {1603.02934},
    archivePrefix = {arXiv},
    primaryClass  = {hep-ex}
}

@techreport{ATLAS:2016xha,
    author        = {{ATLAS collaboration}},
    title         = {{Photon identification in 2015 ATLAS data}},
    institution   = {CERN},
    address       = {Geneva},
    number        = {ATL-PHYS-PUB-2016-014},
    year          = {2016},
    url           = {https://cds.cern.ch/record/2203125}
}

@article{photonidatlas_2019,
    author        = {{ATLAS collaboration}},
    title         = {{Measurement of the photon identification efficiencies with the ATLAS detector using LHC Run 2 data collected in 2015 and 2016}},
    journal       = {Eur. Phys. J. C},
    volume        = {79},
    pages         = {205},
    year          = {2019},
    doi           = {10.1140/epjc/s10052-019-6650-6},
    eprint        = {1810.05087},
    archivePrefix = {arXiv},
    primaryClass  = {hep-ex}
}

@article{electron_id,
    author        = {{ATLAS collaboration}},
    title         = {{Electron reconstruction and identification in the ATLAS experiment using the 2015 and 2016 LHC proton--proton collision data at $\sqrt{s} = 13\,\text{TeV}$}},
    journal       = {Eur. Phys. J. C},
    volume        = {79},
    pages         = {639},
    year          = {2019},
    doi           = {10.1140/epjc/s10052-019-7140-6},
    eprint        = {1902.04655},
    archivePrefix = {arXiv},
    primaryClass  = {hep-ex}
}

@article{egamma_energy_calib,
    author        = {{ATLAS collaboration}},
    title         = {{Electron and photon energy calibration with the ATLAS detector using LHC Run 2 data}},
    journal       = {JINST},
    volume        = {19},
    pages         = {P02009},
    year          = {2024},
    doi           = {10.1088/1748-0221/19/02/P02009},
    eprint        = {2309.05471},
    archivePrefix = {arXiv},
    primaryClass  = {hep-ex}
}

@techreport{ATL-PHYS-PUB-2022-022,
    author        = {{ATLAS collaboration}},
    title         = {{Identification of electrons using a deep neural network in the ATLAS experiment}},
    institution   = {CERN},
    address       = {Geneva},
    number        = {ATL-PHYS-PUB-2022-022},
    year          = {2022},
    url           = {https://cds.cern.ch/record/2809283}
}

@techreport{ATL-PHYS-PUB-2023-001,
    author        = {{ATLAS collaboration}},
    title         = {{Electron Identification with a Convolutional Neural Network in the ATLAS Experiment}},
    institution   = {CERN},
    address       = {Geneva},
    number        = {ATL-PHYS-PUB-2023-001},
    year          = {2023},
    url           = {https://cds.cern.ch/record/2850666}
}

@article{photonidcms_2015,
    author        = {{CMS collaboration}},
    title         = {{Performance of photon reconstruction and identification with the CMS detector in proton-proton collisions at $\sqrt{s} = 8\,\text{TeV}$}},
    journal       = {JINST},
    volume        = {10},
    pages         = {P08010},
    year          = {2015},
    doi           = {10.1088/1748-0221/10/08/P08010},
    eprint        = {1502.02702},
    archivePrefix = {arXiv},
    primaryClass  = {physics.ins-det}
}

@article{Sirunyan_2021,
    author        = {{CMS collaboration}},
    title         = {{Electron and photon reconstruction and identification with the CMS experiment at the CERN LHC}},
    journal       = {JINST},
    volume        = {16},
    pages         = {P05014},
    year          = {2021},
    doi           = {10.1088/1748-0221/16/05/P05014}
}

@article{deOliveira:2018hva,
    author        = {de Oliveira, Luke and Nachman, Benjamin and Paganini, Michela},
    title         = {{Electromagnetic Showers Beyond Shower Shapes}},
    journal       = {Nucl. Instrum. Meth. A},
    volume        = {951},
    pages         = {162879},
    year          = {2020},
    doi           = {10.1016/j.nima.2019.162879},
    eprint        = {1806.05667},
    archivePrefix = {arXiv},
    primaryClass  = {hep-ex}
}

@article{Komiske_2019,
    author        = {Komiske, Patrick T. and Metodiev, Eric M. and Thaler, Jesse},
    title         = {{Energy flow networks: deep sets for particle jets}},
    journal       = {JHEP},
    volume        = {01},
    pages         = {121},
    year          = {2019},
    doi           = {10.1007/JHEP01(2019)121},
    eprint        = {1810.05165},
    archivePrefix = {arXiv},
    primaryClass  = {hep-ph}
}

@techreport{ATL-PHYS-PUB-2022-027,
    author        = {{ATLAS collaboration}},
    title         = {{Graph Neural Network Jet Flavour Tagging with the ATLAS Detector}},
    institution   = {CERN},
    address       = {Geneva},
    number        = {ATL-PHYS-PUB-2022-027},
    year          = {2022},
    url           = {https://cds.cern.ch/record/2811135}
}

@article{atlascollaboration2025transformingjetflavourtagging,
    author        = {{ATLAS collaboration}},
    title         = {{Transforming jet flavour tagging at ATLAS}},
    year          = {2025},
    eprint        = {2505.19689},
    archivePrefix = {arXiv},
    primaryClass  = {hep-ex}
}

@article{Sun:2930164,
    author        = {Sun, Chang and Ngadiuba, Jennifer and Pierini, Maurizio and Spiropulu, Maria},
    title         = {{Fast Jet Tagging with MLP-Mixers on FPGAs}},
    journal       = {Mach. Learn. Sci. Tech.},
    volume        = {6},
    number        = {3},
    pages         = {035025},
    year          = {2025},
    doi           = {10.1088/2632-2153/adf596},
    eprint        = {2503.03103},
    archivePrefix = {arXiv},
    reportNumber  = {FERMILAB-PUB-25-0180-PPD}
}

@article{Barr:2025djz,
    author        = {Barr, Jackson and others},
    title         = {{Salt: Multimodal Multitask Machine Learning for High Energy Physics}},
    journal       = {J. Open Source Softw.},
    volume        = {10},
    number        = {112},
    pages         = {7217},
    year          = {2025},
    doi           = {10.21105/joss.07217}
}

@misc{atlas_salt_docs_2025,
    title         = {{SALT documentation}},
    howpublished  = {\url{https://ftag-salt.docs.cern.ch/}},
    year          = {2025},
    note          = {Accessed: 2025-08-15}
}

@article{Simkina:2023bjm,
    author        = {Simkina, Polina and Couderc, Fabrice and Malcl{\`e}s, Julie and Sahin, Mehmet {\"O}zg{\"u}r},
    title         = {{Reconstruction of electromagnetic showers in calorimeters using Deep Learning}},
    journal       = {Eur. Phys. J. C},
    volume        = {84},
    pages         = {639},
    year          = {2024},
    doi           = {10.1140/epjc/s10052-024-12978-1},
    eprint        = {2311.17914},
    archivePrefix = {arXiv},
    primaryClass  = {hep-ex}
}

@article{Maidannyk_2026,
    author        = {Maidannyk, Yuliia and Couderc, Fabrice and Malcl{\`e}s, Julie and Sahin, Mehmet {\"O}zg{\"u}r},
    title         = {{Reconstruction of overlapping electromagnetic showers in calorimeters using Transformers}},
    journal       = {Eur. Phys. J. C},
    volume        = {86},
    pages         = {873},
    year          = {2026},
    doi           = {10.1140/epjc/s10052-026-16097-x},
    eprint        = {2603.18172},
    archivePrefix = {arXiv},
    primaryClass  = {hep-ex}
}

@techreport{CMS-DP-2026-135,
    author        = {{CMS collaboration}},
    title         = {{Transformer-based reconstruction of electromagnetic showers}},
    institution   = {CERN},
    address       = {Geneva},
    type          = {CMS Detector Performance Summary},
    number        = {CMS-DP-2026/135},
    year          = {2026},
    url           = {https://cds.cern.ch/record/2972650}
}

@inproceedings{vaswani2023attentionneed,
    author        = {Vaswani, Ashish and Shazeer, Noam and Parmar, Niki and Uszkoreit, Jakob and Jones, Llion and Gomez, Aidan N. and Kaiser, Lukasz and Polosukhin, Illia},
    title         = {{Attention Is All You Need}},
    booktitle     = {Advances in Neural Information Processing Systems},
    volume        = {30},
    year          = {2017},
    eprint        = {1706.03762},
    archivePrefix = {arXiv},
    primaryClass  = {cs.CL}
}

@inproceedings{xiong2020layernormalizationtransformerarchitecture,
    author        = {Xiong, Ruibin and Yang, Yunchang and He, Di and Zheng, Kai and Zheng, Shuxin and Xing, Chen and Zhang, Huishuai and Lan, Yanyan and Wang, Liwei and Liu, Tie-Yan},
    title         = {{On Layer Normalization in the Transformer Architecture}},
    booktitle     = {Proceedings of the 37th International Conference on Machine Learning (ICML)},
    year          = {2020},
    eprint        = {2002.04745},
    archivePrefix = {arXiv},
    primaryClass  = {cs.LG}
}

@inproceedings{tolstikhin2021mlpmixerallmlparchitecturevision,
    author        = {Tolstikhin, Ilya and Houlsby, Neil and Kolesnikov, Alexander and Beyer, Lucas and Zhai, Xiaohua and Unterthiner, Thomas and Yung, Jessica and Steiner, Andreas and Keysers, Daniel and Uszkoreit, Jakob and Lucic, Mario and Dosovitskiy, Alexey},
    title         = {{MLP-Mixer: An all-MLP Architecture for Vision}},
    booktitle     = {Advances in Neural Information Processing Systems},
    volume        = {34},
    year          = {2021},
    eprint        = {2105.01601},
    archivePrefix = {arXiv},
    primaryClass  = {cs.CV}
}

@inproceedings{li2017gatedgraphsequenceneural,
    author        = {Li, Yujia and Tarlow, Daniel and Brockschmidt, Marc and Zemel, Richard},
    title         = {{Gated Graph Sequence Neural Networks}},
    booktitle     = {International Conference on Learning Representations (ICLR)},
    year          = {2016},
    eprint        = {1511.05493},
    archivePrefix = {arXiv},
    primaryClass  = {cs.LG}
}

@article{SubmanifoldSparseConvNet,
    author        = {Graham, Benjamin and van der Maaten, Laurens},
    title         = {{Submanifold Sparse Convolutional Networks}},
    year          = {2017},
    eprint        = {1706.01307},
    archivePrefix = {arXiv},
    primaryClass  = {cs.NE}
}

@article{elfwing2017sigmoidweightedlinearunitsneural,
    author        = {Elfwing, Stefan and Uchibe, Eiji and Doya, Kenji},
    title         = {{Sigmoid-Weighted Linear Units for Neural Network Function Approximation in Reinforcement Learning}},
    journal       = {Neural Netw.},
    volume        = {107},
    pages         = {3},
    year          = {2018},
    doi           = {10.1016/j.neunet.2017.12.012},
    eprint        = {1702.03118},
    archivePrefix = {arXiv},
    primaryClass  = {cs.LG}
}

@inproceedings{kingma2017adammethodstochasticoptimization,
    author        = {Kingma, Diederik P. and Ba, Jimmy},
    title         = {{Adam: A Method for Stochastic Optimization}},
    booktitle     = {International Conference on Learning Representations (ICLR)},
    year          = {2015},
    eprint        = {1412.6980},
    archivePrefix = {arXiv},
    primaryClass  = {cs.LG}
}

@inproceedings{loshchilov2019decoupledweightdecayregularization,
    author        = {Loshchilov, Ilya and Hutter, Frank},
    title         = {{Decoupled Weight Decay Regularization}},
    booktitle     = {International Conference on Learning Representations (ICLR)},
    year          = {2019},
    eprint        = {1711.05101},
    archivePrefix = {arXiv},
    primaryClass  = {cs.LG}
}

@article{saleh2024statisticalpropertieslogcoshloss,
    author        = {Saleh, Resve A. and Saleh, A. K. Md. Ehsanes},
    title         = {{Statistical Properties of the log-cosh Loss Function Used in Machine Learning}},
    year          = {2022},
    eprint        = {2208.04564},
    archivePrefix = {arXiv},
    primaryClass  = {stat.ML}
}

@inproceedings{Chen_2016,
    author        = {Chen, Tianqi and Guestrin, Carlos},
    title         = {{XGBoost: A Scalable Tree Boosting System}},
    booktitle     = {Proceedings of the 22nd ACM SIGKDD International Conference on Knowledge Discovery and Data Mining},
    pages         = {785},
    year          = {2016},
    doi           = {10.1145/2939672.2939785},
    eprint        = {1603.02754},
    archivePrefix = {arXiv},
    primaryClass  = {cs.LG}
}

@inproceedings{pytorch,
    author        = {Paszke, Adam and Gross, Sam and Massa, Francisco and Lerer, Adam and Bradbury, James and Chanan, Gregory and Killeen, Trevor and Lin, Zeming and Gimelshein, Natalia and Antiga, Luca and Desmaison, Alban and K{\"o}pf, Andreas and Yang, Edward and DeVito, Zach and Raison, Martin and Tejani, Alykhan and Chilamkurthy, Sasank and Steiner, Benoit and Fang, Lu and Bai, Junjie and Chintala, Soumith},
    title         = {{PyTorch: An Imperative Style, High-Performance Deep Learning Library}},
    booktitle     = {Advances in Neural Information Processing Systems},
    volume        = {32},
    year          = {2019},
    eprint        = {1912.01703},
    archivePrefix = {arXiv},
    primaryClass  = {cs.LG}
}

@inproceedings{tensorflow,
    author        = {Abadi, Mart{\'i}n and Barham, Paul and Chen, Jianmin and Chen, Zhifeng and Davis, Andy and Dean, Jeffrey and Devin, Matthieu and Ghemawat, Sanjay and Irving, Geoffrey and Isard, Michael and Kudlur, Manjunath and Levenberg, Josh and Monga, Rajat and Moore, Sherry and Murray, Derek G. and Steiner, Benoit and Tucker, Paul and Vasudevan, Vijay and Warden, Pete and Wicke, Martin and Yu, Yuan and Zheng, Xiaoqiang},
    title         = {{TensorFlow: A system for large-scale machine learning}},
    booktitle     = {12th USENIX Symposium on Operating Systems Design and Implementation (OSDI 16)},
    pages         = {265},
    year          = {2016},
    eprint        = {1605.08695},
    archivePrefix = {arXiv},
    primaryClass  = {cs.DC}
}

@misc{chollet2015keras,
    author        = {Chollet, Fran{\c{c}}ois and others},
    title         = {{Keras}},
    howpublished  = {\url{https://keras.io}},
    year          = {2015}
}

@misc{spconv2022,
    author        = {{Spconv Contributors}},
    title         = {{Spconv: Spatially Sparse Convolution Library}},
    howpublished  = {\url{https://github.com/traveller59/spconv}},
    year          = {2022}
}
